\documentstyle[psfig,amsmath,epsfig,subfigure]{mn}

\newcommand{\bOmega}{\mbox{\boldmath $\Omega$}}

\newcommand{\blde}{\mbox{\boldmath $e$}}
\newcommand{\bg}{\mbox{\boldmath $g$}}

\newcommand{\bj}{\mbox{\boldmath $j$}}

\newcommand{\bu}{\mbox{\boldmath $u$}}

\newcommand{\bB}{\mbox{\boldmath $B$}}

\newcommand{\ptl}{\partial}

\def\b0{\mbox{ {\bf 0}}}
\def\ltsima{\mbox{$\; \buildrel < \over \sim \;$}}
\def\simlt{\lower.5ex\hbox{\ltsima}}
\def\gtsima{\mbox{$\; \buildrel > \over \sim \;$}}
\def\simgt{\lower.5ex\hbox{\gtsima}}
\def\div{{\mathbf \nabla \cdot}}
\def\curl{{\mathbf \nabla \times}}
\def\grad{{\mathbf \nabla}}

\def\er{{ \hat{\blde}_r}}
\def\etheta{{ \hat{\blde}_{\theta}}}

\def\ez{{ \hat{\blde}_z}}

\def\rc{r_{\rm c}}

\def\rin{r_{\rm in}}
\def\rout{r_{\rm out}}

\def\ur{u_r}
\def\ut{u_{\theta}}
\def\up{u_{\phi}}
\def\Br{B_r}
\def\Bt{B_{\theta}}
\def\Bp{B_{\phi}}

\def\st{\sin\theta}

\def\s2t{\sin^2\theta}
\def\c2t{\cos^2\theta}
\def\sin{\mbox{  } {\rm sin}}
\def\cos{\mbox{  }{\rm cos}}

\def\Oc{\Omega_{\rm c}}
\def\Oeq{\Omega_{\rm eq}}
\def\Oin{\Omega_{\rm in}}

\def\dd{{\rm d}}

\def\Eeta{E_{\eta}}
\def\Enu{E_{\nu}}

\def\Ot{\tilde{\Omega}}

\title{Dynamics of the solar tachocline I: an incompressible study}
\author[P. Garaud]{P. Garaud, \\  Department of Applied Mathematics and Theoretical Physics, University of Cambridge, Silver Street, CB39EW Cambridge, UK \\ Institute of Astronomy, University of Cambridge, Madingley Road, CB30HA Cambridge, UK \\ New Hall, Huntingdon Road, CB30DF Cambridge, UK}

\begin{document}
\maketitle

\begin{abstract}

Gough \& McIntyre have suggested that the dynamics of the solar tachocline are
dominated by the advection-diffusion balance between the differential
rotation, a large-scale primordial field and baroclinicly driven
meridional motions. This paper presents the first part of a study of
the tachocline, in which a model of the rotation profile
below the convection zone is constructed along the lines suggested by
Gough \& McIntyre and solved numerically. In this first part, a
reduced model of the tachocline is derived in which the effects of
compressibility and energy transport on the system are neglected; the
meridional motions are driven instead by Ekman-Hartmann
pumping. Through this simplification, the interaction of the fluid
flow and the magnetic field can be isolated and is studied through
nonlinear numerical analysis for
various field strengths and diffusivities. It is shown that there
exists only a narrow range of magnetic field strengths for which the
system can achieve a nearly uniform rotation. The results are discussed with
respect to observations and to the limitations of this initial
approach. A following paper combines the effects of realistic
baroclinic driving and stratification with a model that follows
closely the lines of work of Gough \& McIntyre.

\end{abstract}
%\begin{keywords}
%MHD -- Sun:magnetic fields -- Sun:interior -- Sun:rotation
%\end{keywords}

\section{Introduction}

The solar tachocline is a thin shear layer located at the top of the
radiative zone. It performs the dynamical transition between the
differentially rotating convection zone and the nearly uniformly
rotating radiative zone (Brown et al., 1989). In a non-rotating fluid, the interface between a convective and a non-convective region is intrinsically
complex, and supports dynamical phenomena such as
overshoot and gravity-wave generation. Including a background rotation
increases the complexity of the system by driving
meridional motions at the interface (Mestel, 1953) and by changing
the characteristics of convective motions and overshooting plumes, both
leading in particular to non-vanishing anisotropic stresses
and differential rotation. Finally, the combination of strong shear
and anisotropic small-scale motion has been shown to lead sometimes to the
enhancement of any seed magnetic field, and to intermittent 
(or eventually quasi-periodic) magnetic phenomena which could be
related to the observed 11-yr magnetic cycle. The position of the
tachocline in this unique dynamical interface makes it one of the most
complex, and as a result, one of the most interesting regions of the
sun.

In recent years, the tachocline has been one of the principal
targets of helioseismic observations of the angular rotation profile
of the sun. The determination of the thickness of the tachocline is 
essential to understanding its dynamics.  Assuming a variation 
profile for the angular velocity
across the tachocline, throughout the radiative interior and the convection
zone, artificial frequency splittings can be reconstructed which are
then fitted to the observations. The most recent estimate of the
thickness of the tachocline $\Delta$ using this method is the following: 
$\Delta = (0.039 \pm 0.013) {\rm R}_{\odot}$ (Charbonneau et al.,1999). 
An independent method is proposed by Elliott \& Gough
(1999): they suggest that the discrepancy between the observed sound-speed 
profile and that obtained from Standard Solar Models could be
related to a chemical composition anomaly caused by the mixing of helium
within the tachocline layer. By including an additional mixing layer
below the convection zone, and comparing the predicted sound speed of this new model to the observations, they calibrate the thickness of the 
tachocline to be approximately 0.02 ${\rm R}_{\odot}$, assuming the tachocline to be strictly spherical. If the tachocline is thinner but aspherical, this estimate is a measure of the asphericity. Note that observations of light-elements abundances at the surface also suggest that mixing below the convection zone is confined to a shallow layer (e.g. Brun, Turck-Chi\`eze \& Zahn, 1999).

It has also been suggested that helioseismic observations could
be used to determine the latitudinal structure of the tachocline,
in particular whether the width and position of the tachocline 
vary with latitude.
 Independent analyses by Gough \& Kosovichev (1995) and
Charbonneau et al. (1999) suggest that the base of the convection zone
and the tachocline (respectively) may be prolate. Charbonneau et
al. (1999) also report that no significant variation in the tachocline
thickness can be deduced from the observations. Finally, the most
recent observational results on the structure of the tachocline
concern its dynamical aspect: Howe et al. (2000) suggest that a
large-scale oscillation may be taking place across the tachocline with
a period of about 1.3 yr, although these results are contested 
(Basu \& Antia, 2001). If assumed to be related to Alfv\'enic
torsional oscillations, this observation could be evidence for the
presence of a magnetic field threading the tachocline with a radial
component of about 500 G (Gough, 2000).

Owing to the complexity of the dynamics of this region, only a handful
of models of the tachocline have been proposed so far. Spiegel \&
Zahn (1992) pointed out that the shear across the tachocline must be
associated with significant thermal fluctuations through a thermal-wind 
balance. These temperature fluctuations drive meridional
motions and propagate the shear deeper and deeper into the
radiative zone. In order to avoid this radiative spreading, Spiegel
\& Zahn proposed an anisotropic turbulence stress-model to suppress
the latitudinal shear within a short vertical lengthscale, 
the strong stratification of
the radiative zone providing a natural explanation for the anisotropy
of the small-scale flow. However, it was later argued that such a
model cannot adequately describe the tachocline (Gough, 1997, Gough \&
McIntyre, 1998). Broadly speaking, the strong restriction of turbulent
 motions to spherical shells by the background
stratification in the radiative zone leads to Reynolds stresses that
transport not angular momentum but potential vorticity (see Garaud,
2001a, for instance). These Reynolds stresses would therefore drive
the system away from rather than towards uniform rotation. Moreover,
Spiegel \& Zahn's model of the tachocline assumes  that the tachocline
is turbulent; it is not clear whether this is indeed the
case. The tachocline is likely to be stable (Charbonneau, Dikpati \& Gilman, 1999) or marginally stable (Garaud, 2001a) to hydrodynamical linear
shear instabilities. However, by assuming the
existence of a relatively strong toroidal field, it was shown that MHD
2-D instabilities exist (Dikpati \& Gilman, 1999) and can have a significant influence on the redistribution of angular momentum (Cally, 2001). 

One of the most natural models for the solar interior rotation, 
and in particular the
quenching of the shear, is to assume the existence of a large-scale
primordial field in the radiative zone. Indeed, as the magnetic
diffusion timescale in the radiative zone is much larger than the
rotation timescale, Ferraro's isorotation law (1937) ensures that the
angular velocity is constant on field lines. Mestel \& Weiss (1987)
suggested that large-scale fields of amplitude as low as
$10^{-3}-10^{-2}$G should be capable of suppressing most of the
rotational shear deep in the solar radiative interior. Simulations
performed by R\"udiger \& Kitchatinov (1997) and MacGregor \&
Charbonneau (1999) looked at the interaction between a large-scale field
with a fixed poloidal component and a purely azimuthal flow. These
seemed to confirm this estimate, and indeed managed to reproduce the
confinement of the shear to a thin tachocline. However, as it was
pointed out by Gough \& McIntyre (1998), the very low magnetic
diffusivity of the radiative zone enables a strong nonlinear coupling
between the thermally driven meridional flows described by Spiegel \&
Zahn and the large-scale interior field. This interaction is essential 
to the dynamics of the tachocline. Starting from this idea, 
Gough \& McIntyre (1998) presented the first model of the 
tachocline to take into account self-consistently 
the nonlinear interaction between thermally driven meridional 
flows and a large-scale magnetic field. The complexity of the resulting
model, however, precluded the derivation of a complete solution.

It is the purpose of this work to create a model of the
tachocline which encompasses gradually more realistic physics, to reproduce
 eventually the idea proposed by Gough \& McIntyre. Only this type of 
gradual approach can provide an adequate 
procedure to isolate different 
dynamical phenomena and to study them individually 
before combining them into one complex system. 
In this first paper, the compressibility of the fluid is
neglected (as well as the background stratification and the energy
transport), in order to study specifically the nonlinear interaction
between the magnetic field and the fluid motions. As a result,
meridional motions must be driven artificially. This can be done through 
Ekman-Hartmann
pumping on the boundary representing the interface with the convection
zone, which has the main advantage of providing a well-known, controlled 
flow pattern. This idea is discussed in detail in Section \ref{sec:discmod}. 
Subsequent
work will study the effects of stratification as well as advective and
diffusive heat transport on the system, which can then provide a
realistic quantitative description of a thermally driven flow and
ultimately of the tachocline.

The initial model studied in this first paper, 
is presented in Section ~\ref{sec:MHDeq}: the assumptions
and boundary conditions are discussed and applied to the system, and
the numerical procedure for the solution of the governing equations is
outlined. In Section \ref{sec:nomagn}, the numerical procedure is
tested by studying a non-magnetic case. The simulations are compared
to the well-known Proudman-Stewartson solution for an incompressible fluid between two
concentric rotating spheres (Proudman, 1956, and Stewartson, 1966). 
The results of the simulations in the magnetic case are presented in
Section \ref{sec:magn}. It is shown that the effects of angular-momentum
advection and poloidal-field advection by the meridional flow are essential to
the dynamics of the system, thereby emphasizing the importance of correctly taking both into account in realistic models of the tachocline. In particular it is shown that, contrary to common expectations, increasing the amplitude
of the magnetic field does not necessarily reduce the amount of shear
in the radiative zone. Instead, the simulations presented in Section \ref{sec:magn} reveal the existence
of three typical regimes, with low-, high- and intermediate-field-strength;
only the intermediate regime is able to reproduce the qualitative
features of the tachocline. Section \ref{sec:blanal} presents two boundary-layer analyses which apply locally to different field configurations; the analytical solutions are compared to the numerical simulations. Finally, the simulations are analysed and
discussed in Section \ref{sec:disc} with respect to previous work 
and to observations.

\section{The model}
\label{sec:MHDeq}

The principal aim of this study is to try to reproduce the observed
rotation profile of the sun in the region below the convection zone,
and in particular
\begin{enumerate}
\item the sharp transition (within a thin shear layer of thickness no
larger than 4 \% of the solar radius) between the latitudinal shear
observed at the base of the convection zone and the uniform rotation of the radiative interior.
\item the value of the interior angular velocity $\Oc$ 
(observed to be about 93\% of the surface equatorial angular velocity $\Oeq$). 
\end{enumerate}
Moreover, the model should also aim to reproduce the confinement of the material mixing to the region of the tachocline, as suggested by helioseismic measurements (Elliott \& Gough, 1999) and surface light-elements abundances (Brun, Turck-Chi\`eze \& Zahn, 1999).
 
\subsection{The assumptions}
\label{sec:assump}

The model presented in this paper essentially follows the idea
proposed by Gough \& McIntyre (1998) by assuming the existence of a
large-scale primordial field in the radiative zone and looking at its
nonlinear interaction with fluid motions (both azimuthal and
meridional) near the bottom of the convection zone in particular. In
order to do so, nonlinear MHD equations are solved, in a first
instance assuming that the fluid is incompressible (this paper) and
in subsequent papers in an anelastic approximation of a compressible fluid. 
This gradual increase in the complexity of the phenomena studied
allows the independent study of the nonlinear interaction of the field
and the meridional flow on the one hand, and the effects of heat
transport and stratification on the other hand. The principal shortfall 
of the incompressibility assumption is that the mechanisms driving the 
meridional motions as described by Spiegel \& Zahn (1992) are absent 
from this first model; instead, they are replaced by a judicious choice 
of boundary conditions which lead to Ekman-Hartmann pumping of a 
qualitatively similar flow (see Section \ref{sec:discmod}).
   
Two principal simplifications are performed on the system to reduce it to
a numerically tractable problem: it is assumed that the system is
axisymmetric and is in a steady state. Axisymmetry is a wholly
unjustified, albeit standard assumption, which allows a drastic
simplification of the MHD equations. The steady-state assumption, on
the other hand, is reasonably well justified provided all dynamical
timescales (for the rotation and the meridional circulation) 
are much shorter than the stellar-evolution timescale (which ensures 
that the background stratification
does not vary during the fluid motion), the magnetic-braking timescale
(which ensures that the external torques exerted on the system are small), or
the magnetic-diffusion timescale (which ensures that the amplitude of
the interior field does not vary significantly during the fluid
turnover time). 

In the following, it will always be assumed that the tachocline lies
completely in the stably stratified radiative zone, which agrees
reasonably well with observations (Charbonneau et al. 1999). 
It will also be assumed that the rotation
profile in the convection zone is independent of the tachocline
dynamics so that the convection zone can be taken as a boundary
condition of the model. Finally, it will be assumed that the 
region studied is located sufficiently far below the region of 
influence of the solar dynamo (Garaud, 1999) to avoid complications 
linked with the unsteady character of the dynamo field. As a result 
of this assumption, the region studied is also likely to be located 
below the overshoot region.

Within this framework, the system studied is limited to the radiative
zone, in a region located between two concentric spherical boundaries,
the outer one near the base of the convection zone $\rout=\rc$, 
where $\rc = 0.7 {\rm R_{\odot}}$, the
inner one at a radius $\rin = 0.35 \rc$. The inner core (occupying $r<\rin$)
is cut out of the region of the simulation to avoid singularities at
the origin; the position of the inner boundary should not have a
significant influence on the results. The numerical value of $\rin$ is 
chosen in such way as to facilitate the comparison of these simulations 
with the work of Dormy, Cardin \& Jault (1998), who solved a similar 
system of equations and boundary conditions for geophysical applications.   

\subsection{The equations}
\label{sec:eqs}

If the fluid motions are assumed to be axisymmetric and
incompressible, the MHD  equations in a frame rotating with angular
velocity $\bOmega_{\rm c}= \Oc \ez$ are the following:
\begin{eqnarray}
2\rho \bOmega_{\rm c} \times \bu + \grad p + \rho \bg_{\rm h} - \bj
 \times \bB - \rho \nu \grad^2 \bu &=&0  \mbox{   ,} \nonumber \\
\curl(\bu \times \bB) + \eta \grad^2 \bB &=& 0 \mbox{   ,} \nonumber \\
\div \bu &=& 0 \mbox{   ,} \nonumber \\
\div \bB &=& 0 \mbox{   .} 
\end{eqnarray}
where the nonlinear terms in the meridional circulation $(\bu \cdot
\grad) \bu$  are neglected (it is verified a posteriori that this is a
good approximation -- see Section \ref{sec:discmod}), and $\bg_{\rm h} = g_{\rm h}(r) \er$ is the
gravitational acceleration associated with the hydrostatic background. 
The angular
velocity $\Omega_{\rm c} = 2.84 \times 10^{-6}$ s$^{-1}$ is chosen to take the
observed interior angular velocity of the sun; $\bB$ is the magnetic
field (with components ($\Br,\Bt,\Bp$)), $\bj$ is the electric current, and $\bu$ is the circulation which has components $(\ur,\ut,\up)$. In order to derive these equations, the density
$\rho $ is assumed to be constant ($\rho_0 =1$g cm$^{-3} $) everywhere, as is
the kinematic viscosity $\nu$ and the magnetic diffusivity $\eta$. This approximation is made to simplify the numerical procedure slightly, but can easily be removed. Since, as it will be shown later, the typical values of the viscosity and magnetic diffusivity used in the simulations are several orders of magnitude larger than in the sun, it seems pointless to try and represent accurately the variation of these quantities in this first analysis. 

Using the following new system of units:
\begin{equation} \left[r\right] = \rc \mbox{   ,   } \left[t\right] =
1/\Oc \mbox{   ,    } \left[B\right] = B_0 \mbox{   ,   } [u]=\rc\Oc \mbox{   ,}
\end{equation} 
where $B_0$ is the typical strength of the radial field in the
interior, the equations become:
\begin{eqnarray}
2 \ez \times \bu &=& - \grad p - \bg_{\rm h}+  \Lambda \bj
\times \bB + \Enu \grad^2 \bu \mbox{   ,} \nonumber \\ \curl(\bu
\times \bB) &=& -\Eeta \grad^2 \bB \mbox{   ,}
\end{eqnarray}
with
\begin{equation}
\Lambda = \frac{ B_0^2}{\rho_0 \Oc^2 \rc^2} \mbox{,    } \Enu =
\frac{\nu}{\Oc \rc^2} \mbox{   and   } \Eeta = \frac{\eta}{\Oc \rc^2}
\mbox{   ,}
\label{eq:elsdef}
\end{equation} 
where all the quantities are now dimensionless and $\ez$ is the unit
vector parallel to the rotation axis. The Ekman numbers $\Enu$ and
$\Eeta$ represent the ratios of the rotation timescale to the diffusive
timescales, and the Elsasser number $\Lambda$ is the ratio of the
typical amplitude of the Lorentz force to that of the Coriolis
force. The system is solved for $(\ur,\ut,\up)$ and $(\Br,\Bt,\Bp)$ by
solving the azimuthal component of the momentum equation, the
azimuthal component of the vorticity equation, the integrated
induction equation and the azimuthal component of the induction
equation, together with the mass conservation equation and the
solenoidal condition; those equations are
\begin{eqnarray}
2 (\ez \times \bu)_{\phi} &=& \Lambda (\bj \times \bB)_{\phi} + \Enu(
\grad^2 \bu)_{\phi}  \nonumber\mbox{   ,} \\ 2 \left[\curl(\ez \times
\bu)\right]_{\phi} &=& \Lambda \left[\curl(\bj \times
\bB)\right]_{\phi} + \Enu( \grad^2 \bomega)_{\phi}  \nonumber\mbox{
,} \\ \ur\Bt - \ut\Br &=& \Eeta  j_{\phi}  \nonumber \mbox{   ,} \\
\left[\curl(\bu \times \bB)\right]_{\phi} &=& -\Eeta (\grad^2
\bB)_{\phi}  \nonumber\mbox{   ,} \\ \div \bu &=& 0 \mbox{   ,}
\nonumber \\ \div \bB &=& 0 \mbox{   ,} 
\label{eq:eqsmhd}
\end{eqnarray}
where $\bomega = \curl \bu$ is the vorticity. 

\subsection{The boundary conditions}
\label{sec:bc}

In this first analysis, the boundary conditions chosen for the system are the simplest
possible ones that still guarantee the existence of a solution; as a
result, they are not necessarily the most accurate representation of
the dynamics of the sun (and in particular of the interface of the
tachocline with the convection zone). The
effects of the boundary conditions on the system, and possible
improvements, are discussed in Section \ref{sec:discmod}

\begin{figure}
\epsfig{file=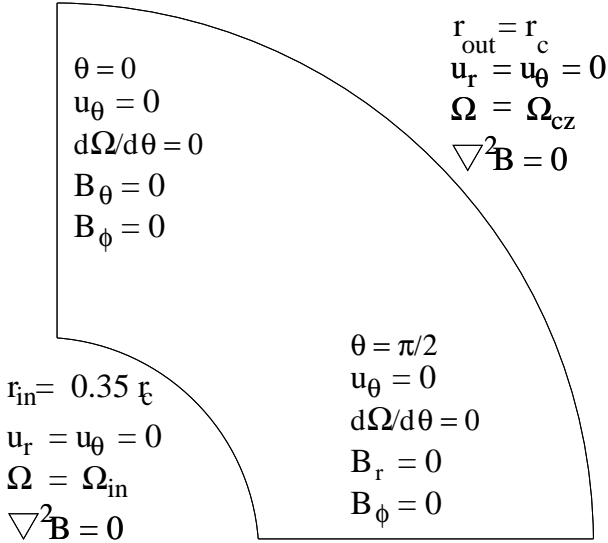,width=8cm}
\caption{\small Boundary conditions.}
\label{fig:bc}
\end{figure}
The boundary conditions are summarized in Fig. \ref{fig:bc}; 
the MHD equations presented in equations (\ref{eq:eqsmhd}) 
are solved in a region
 located between two spherical boundaries, impermeable and
 no-slip boundaries so that the radial and the latitudinal components of
 the circulation vanish on the boundaries, and the azimuthal component
 of the circulation is given by the rotation of the boundaries. As a
 result, the angular-velocity perturbation is $\tilde{\Omega} =
 \Omega_{\rm in} - \Omega_{\rm c}$ on the inner boundary where
 $\Omega_{\rm in}$ is an eigenvalue of the problem (see below) and
 $\tilde{\Omega} = \Omega_{\rm cz}(\theta) - \Omega_{\rm c}$ on the
 outer boundary, where
\begin{equation}
 \Omega_{\rm cz}(\theta) = \Omega_{\rm eq} (1- a_2\cos^2 \theta - a_4
 \cos^4 \theta) 
\end{equation}
and $\Omega_{\rm eq} = 1.07 \Oc$, $a_2 = a_4 = 0.15$ (according to the
 observations presented by Schou et al. (1998)). On the equatorial plane, symmetry arguments determine the behaviour of the solutions: $\ut = 0$, $\Br = 0$, $\ptl \tilde{\Omega}/\ptl \theta = 0 $ and $\Bp = 0$. On the poles, regularity conditions impose $\ut = 0$, $\Bt = 0$, $\ptl \tilde{\Omega}/\ptl \theta = 0$ and $\Bp = 0$. The bounding spheres
 are assumed to be imperfectly conducting and the region outside
 $[\rin,\rout]$ supports no fluid motion, which implies that the
 field in that region satisfies $\grad^2 \bB =0$, with conditions that
 $\bB \rightarrow 0$ at infinity, and that the field structure 
 becomes purely dipolar
 as $r\rightarrow 0$. The amplitude of the radial component
 of the field near the poles is fixed such that $B_r(\rin,\theta=0) =
 B_0$.  This uniquely determines the boundary conditions for the
 magnetic field, as a relation between $\Br$ and $\Bt$ on the one
 hand, and between $\Bp$ and $\ptl \Bp/\ptl r$ on the other hand. As a
 result of these conditions, when the system is not rotating, the
 solution for the magnetic field is a purely poloidal dipolar
 structure with a field strength varying as $1/r^3$. 
 When the system is rotating, these
 boundary conditions drive a meridional flow through Ekman-Hartmann
 pumping. This flow is used to mimic the baroclinicly driven flow
 predicted to occur in the tachocline (Spiegel \& Zahn, 1992, Gough \&
 McIntyre, 1998), in order to study its interaction with the imposed
 magnetic field. The principal caveat of this approach is that the
 typical velocities of the meridional flow scales with $\Enu$ and
 $\Eeta$ as an Ekman-Hartmann flow rather than as a thermally driven
 flow (i.e. with $\ut \sim \rc\tilde{\Omega}$ and $\ur \sim \delta_{\rm EH} \tilde{\Omega}$ where $\delta_{\rm EH}$ is the Ekman-Hartmann boundary-layer thickness). This discrepancy should influence quantitative predictions
 of the model only whilst leaving qualitative results unaffected.

One of the main aims of this study is to be able to predict the
angular velocity of the interior. In order to do so, the angular velocity 
of the inner core $\Omega_{\rm in}$ is treated as an eigenvalue of the
problem, and an additional boundary condition is imposed on the system
accordingly, namely that no net torque is applied to the
region interior to $r=\rin$ (this is a necessary condition to
guarantee a steady state). This condition, which determines uniquely
the value of $\Omega_{\rm in}$, is equivalent to requiring that the
integral of the angular-momentum flux through the boundary vanishes:
\begin{equation}
\int_{0}^{\pi/2} \left( \rho \nu r^2 \sin^2 \theta \frac{\ptl
\Omega}{\ptl r} + r\sin\theta B_r B_{\phi} \right)  \sin \theta \dd
\theta = 0 \mbox{   .} 
\end{equation}
One can notice through this expression the main justification for
choosing conducting boundary conditions: when the region below $\rin$
is insulating, the radial component of the current
(and thereby the toroidal field) must vanish on the boundary; 
in that case the angular-momentum flux through
the boundary would be purely viscous. Since viscous stresses are normally
thought to be negligible in the sun, such a model would be a poor representation of the dynamical structure of the radiative zone.  

\subsection{Numerical method}
\label{sec:num}

The system of equations (\ref{eq:eqsmhd}) and the boundary conditions
presented in Section \ref{sec:bc} constitute a well-posed partial differential equations system with one
eigenvalue. The numerical method chosen for the solution of this
system is the following:
\begin{enumerate}
\item to write the system of equations with respect to a
spherical polar coordinate system, respecting the symmetry of the
problem as well as the regularity conditions on the poles,
\item to expand the latitudinal dependence of the
equations into Fourier modes of $\theta$, or equivalently Chebishev polynomials $T_n(\cos \theta)$,
\item to solve the resulting ODEs and algebraic equations 
with respect to the independent variable $r$ using a Newton-Raphson relaxation method. 
\end{enumerate}
A more detailed description of the numerical
procedure can be found in the work by Garaud (2001b).

Manual mesh-stretching is preferred over automatic
mesh-point allocation, as the latter was found to have low
performance for boundary-layer thicknesses order of $10^{-6}
(\rout-\rin)$ or less. This low performance is probably due to the
fact that as many Fourier modes are used, the allocation algorithms 
that were tried 
cannot choose adequately according the most relevant function against which
 the mesh should
be stretched. When stretching the mesh manually (according to the
width and positions of the boundary layers determined in Section
\ref{sec:blanal}), a minimum of a hundred points is allocated to each of the
boundary layers. Typically, another 200-400 points are allocated to
the interval outside the boundary layers, depending on the simulations.

\section{Results in the non-magnetic case}
\label{sec:nomagn}

As a first step towards the resolution of the model presented above,
a simplified system is solved in which the magnetic field is ignored.
The hydrodynamics of a fluid between two concentric impermeable
rotating spheres is a relatively well-studied problem, both
analytically (since Proudman, 1956) and numerically (see for instance
Dormy, Cardin \& Jault, 1998); these previous studies can be compared
with the results of the non-magnetic numerical solutions obtained with
the numerical procedure presented in Section \ref{sec:num} to test its
accuracy and performance. 

\subsection{Numerical results}

\begin{figure*}
\centerline{\epsfig{file=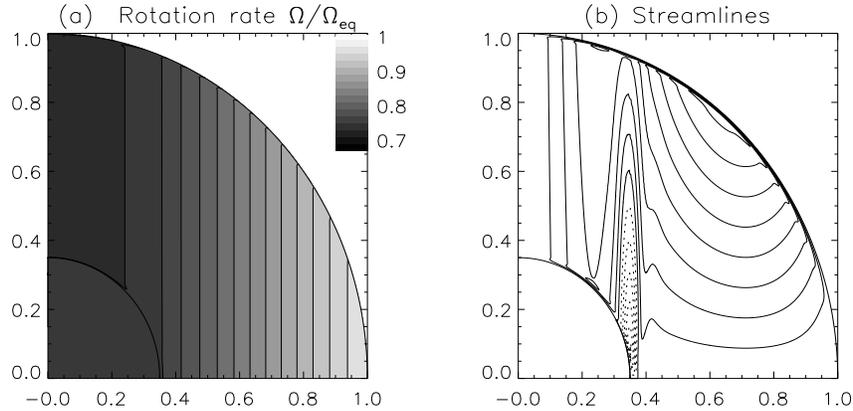,width=12.5cm,height=6cm}}
\caption{\small Rotation profile and streamlines in the non-magnetic
case  for $E_{\nu} = 8 \times 10^{-6}$. The streamlines are line-coded so that clockwise motions are represented by dotted lines and anti-clockwise motions are represented by solid lines. The interior angular velocity in
this  simulation is $\Omega_{\rm in} =  0.75 \Omega_{\rm eq}$. }
\label{fig:nomagn}   
\end{figure*}

As a example, the solution to the non-magnetic subset of equations (\ref{eq:eqsmhd}) with an Ekman number $\Enu = 8 \times 10^{-6}$ is presented in Fig. \ref{fig:nomagn}.
When no magnetic field is present, for low enough Ekman number, fluid motion is dominated by
Coriolis forces everywhere except in two boundary layers near the
spherical boundaries, and in a shear layer at the tangent cylinder. In
the bulk of the fluid, angular  velocity is more-or-less constant on
cylinders: indeed, when viscosity is negligible, the fluid dynamics
equations for the incompressible fluid reduce to
\begin{equation}
(\bOmega_{\rm c} \times \bu)_{\rm \phi} = 0 \mbox{   ,}
\end{equation}
which implies that $\bu$ must be parallel to the rotation axis, and
\begin{equation}
\left(\curl (\bOmega_{\rm c} \times \bu)\right)_{\rm \phi} = 0 \mbox{   ,}
\end{equation}
which implies that the angular velocity must be independent of $z$
where $z$ is the cylindrical coordinate that runs parallel to the
rotation axis (Proudman, 1916, Taylor, 1921).  Viscous effects are necessary in the boundary layers
to ensure the smooth transition between the rotation profile in the
bulk of the fluid and that imposed at the boundaries.

\subsection{Comparison with analytical asymptotic analysis}
\label{sec:nomagn.anal}

\begin{figure}
\centerline{\epsfig{file=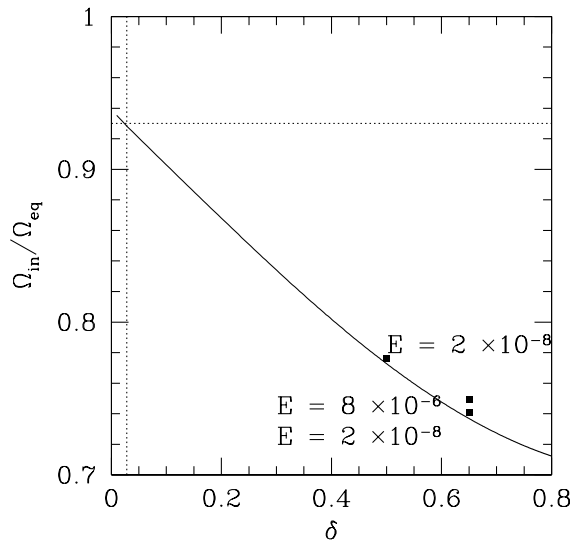,width=8cm}}
\caption{\small Analytical prediction for the interior angular velocity
as a function of gap width $\delta$, and results of simulations for
$\delta = 0.5$ and $\delta = 0.65$ for values of the Ekman number as
shown.}
\label{fig:ocdelta}
\end{figure}

 This structure was first studied by Proudman (1956) and Stewartson
 (1966), in the case where $E_{\nu}$ is asymptotically small. It is
 possible to show (see the Appendix) that in this limit the interior
 angular velocity is uniquely determined by the size of the gap
 between the two spheres; its value can be predicted analytically, and
 is given by the following equation:
\begin{equation}
\frac{\Omega_{\rm in}}{\Omega_{\rm eq} } = \frac{\int_0^{\pi/2}
F(\theta) D(\sin \theta) \dd \theta}{\int_0^{\pi/2} F(\theta) \dd
\theta}
\label{eq:omA} \mbox{   ,}
\end{equation}
where 
\begin{equation}
F(\theta) = \frac{\sin^3\theta  \cos\theta}{\cos^{1/2}\theta +
\left(1-\s2t \rin^2/\rout^2 \right)^{1/4}} 
\end{equation}
and
\begin{equation}
D(s) = 1-a_2
\left(1-(s^2/\rout^2)\right)-a_4\left(1-(s^2/\rout^2)\right)^2 \mbox{   .}
\label{eq:ds} 
\end{equation}
The variation of the calculated value of the interior anguar velocity as
a function of the width of the gap $\delta = \rout-\rin$ is presented
in Fig ~\ref{fig:ocdelta}.
It can be seen that the simulations, represented by the black squares,
fit well the analytical predictions provided that the Ekman number is small
enough (i.e. below $10^{-6}$). It is also interesting to note, as an
aside, that for gap width of about 3\% of the radiative zone's radius
(which corresponds to the width of the solar tachocline), the interior
angular velocity is 93\% of the equatorial velocity, which is very
close to the value observed. It is not
clear whether this interesting match is a mere fortuitous 
coincidence or the result of some more subtle physical processes.

\begin{figure}
\centerline{\epsfig{file=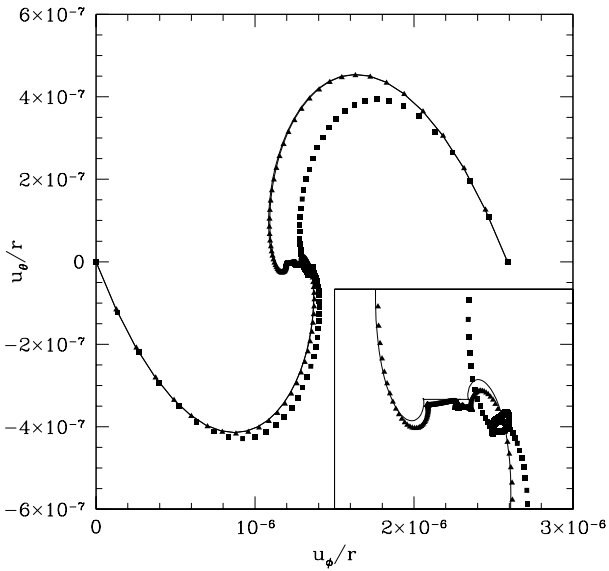,width=8cm}}
\caption{\small Predicted and calculated Ekman spirals. The analytical
prediction in shown  as a continuous line, whereas the simulations are
shown as squares for $\Enu=10^{-4}$ or triangles for
$\Enu = 6.5 \times 10^{-6}$.  The inset shows an enlargement of the
central area.}
\label{fig:ekmspir}
\end{figure}

Another way of comparing the results of the simulations to analytical
predictions is through the construction of the Ekman spiral, which is
a parametric representation of the azimuthal velocity against the
latitudinal velocity as a function of radius, at a fixed co-latitude
$\theta$. Fig. \ref{fig:ekmspir} compares the results of the
asymptotic solution and the true numerical solution for a slightly
different simulation, in which the angular-velocity profile imposed
on the outer boundary is chosen to be constant with value $\Omega_{\rm
out} = \Omega_{\rm c} + 10^{-5}$, and the angular velocity of the
inner core is simply $\Omega_{\rm in} = \Omega_{\rm c}$ (the no-torque
condition is dropped).
Note how the fit of the asymptotic analytical prediction to the
numerical solution is valid only provided the Ekman number is small
enough. For larger Ekman numbers, viscosity plays a non-negligible
role in the dynamics of the fluid outside the boundary layers,
invalidating Proudman's asymptotic analysis. This result has been
obtained already by Dormy, Cardin \& Jault (1998) who studied the non-magnetic
case extensively. The good agreement between the analytical asymptotic solutions and the simulations validates the numerical procedure.

\section{Results in the magnetic case}
\label{sec:magn}

The influence of the magnetic field on the fluid
depends essentially on two parameters: the field strength and the
magnetic diffusivity. In this section, three regimes are
presented for varying Elsasser number at fixed $(\Enu,\Eeta)$. The Elsasser 
number is then fixed, and in Section \ref{sec:intfield2} the 
dependence of the solution on the magnetic Ekman number is presented.

\subsection{Varying the field strength}

In these first simulations, only the Elsasser number $\Lambda$ is
varied. Note that the definition of $\Lambda$ defined in equation
(\ref{eq:elsdef}) uses the value of the amplitude of the radial
component of the magnetic field on the inner boundary. Accordingly, it
should normally be defined using the true value of the density at $r=\rin$, 
which in reality is of order of $\rho_{\rm in} = 20$ g cm$^{-3}$
rather than the chosen uniform value of $\rho_0 = 1$ g cm$^{-3}$. As a result
of the uniform-density approximation, the values of $\Lambda$ chosen in 
this work should be interpreted with care and regarded as rough
indications rather than precise values. Note also that as the
``initial'' magnetic field varies as $1/r^3$, the typical local
Elsasser number near the surface is about 500 times lower than $\Lambda$.

The values of the viscous and magnetic Ekman numbers chosen 
for these simulations
are identical: $\Enu = \Eeta = 2.5 \times
10^{-4}$. This value was chosen for convinience, as it is then easy to
obtain solutions for any value of the magnetic field strength.

\subsubsection{Low-field case, $\Lambda = 1/25$}
\label{sec:lofield}

\begin{figure*}
\centerline{\epsfig{file=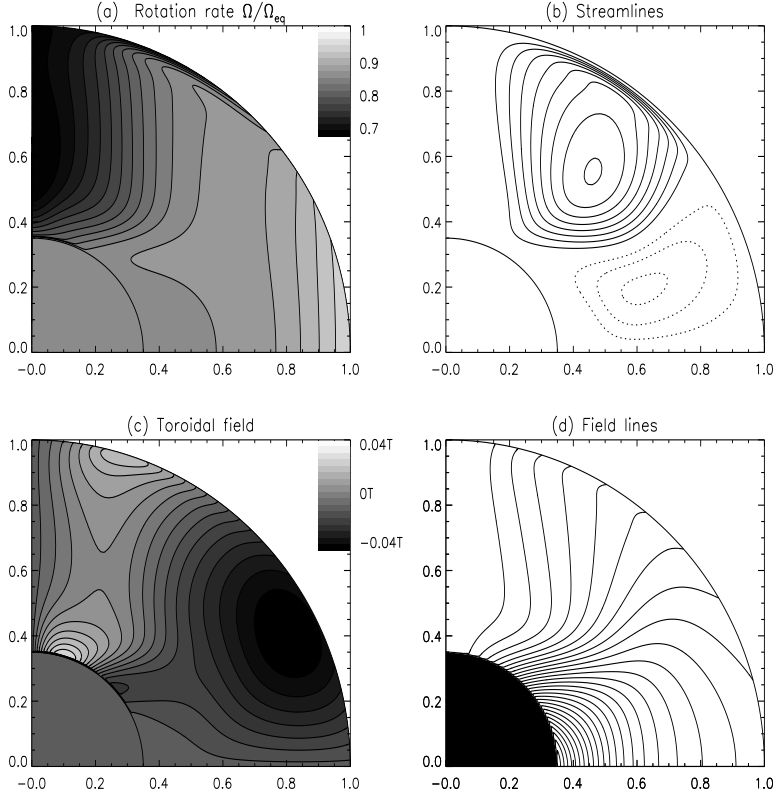,height=11cm,width=11cm}}
\vspace{-0.2cm}
\caption{\small Simulation results for $\Lambda = 1/25$, $E_{\nu} =
2.5\times 10^{-4}$, and $E_{\eta} = 2.5\times 10^{-4}$. Same line-style coding as in Fig. \ref{fig:nomagn} for the streamlines.}
\label{fig:lowfield}
\end{figure*}

This first simulation is shown in Fig. \ref{fig:lowfield}, which
presents the result in the case of a low Elsasser number ($\Lambda =
1/25$).  This corresponds to $B_0 = 0.25$ T. The local Elsasser number
near the surface is of order of $7 \times 10^{-5}$.  

The structure of the interior angular velocity is dominated by
Coriolis forces, and the angular-velocity profile is close to Proudman
(cylindrical) rotation, except maybe near the inner core where the
influence of a magnetic field can be seen through the slight deviation
in the angular-velocity contour lines. Because of the additional
Lorentz forces in the momentum equation, the circulation is no more
limited to cylindrical surfaces and takes a rather different pattern,
with two cells that burrow deeply into the radiative zone. As
expected from standard Ekman-Hartmann pumping (see Acheson \& Hide,
1973), the typical value of the latitudinal component of the flow near
the top boundary is of order of $r \tilde{\Omega}$, i.e. $\ut \sim
1.5 \times 10^{-2}$ in units of $\rc\Oc$. 
The advection of the poloidal field by the
circulation can be seen in Fig. \ref{fig:lowfield}(d) where the field
lines are dragged equatorwards in the interior and polewards near the
top boundary by a strong anticlockwise polar cell. The shear persists
throughout the fluid region, and as a result, leads to the winding up
of the poloidal field into a relatively strong toroidal field. Typical
values of the toroidal field are of order of one tenth of the value of
the poloidal field near the core. This structure shows little
resemblance with the observations, failing in particular to impose
uniform rotation within the fluid region.

\subsubsection{High-field case, $\Lambda = 25$}
\label{sec:hifield}

\begin{figure*}
\centerline{\epsfig{file=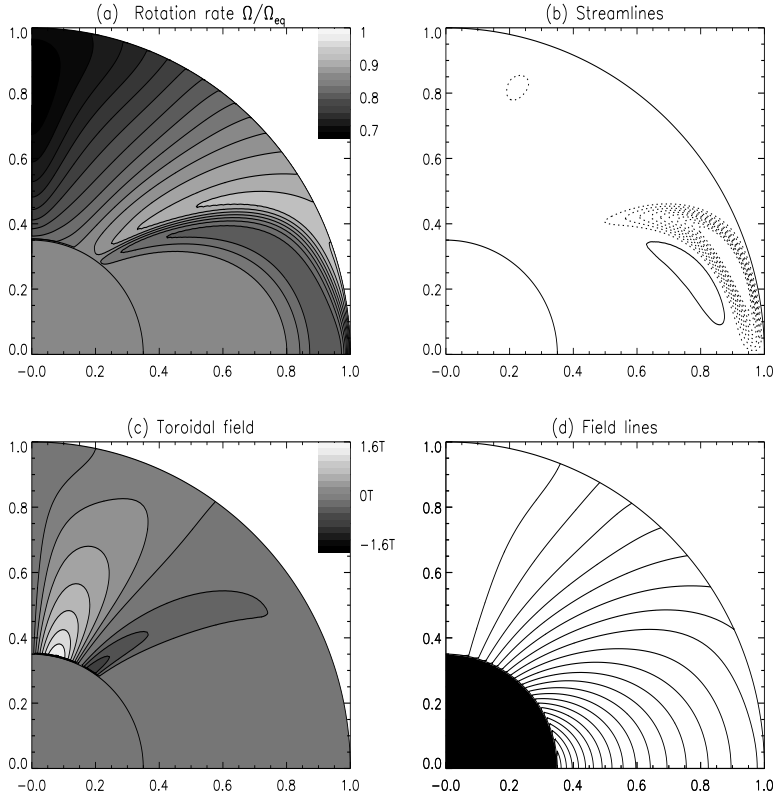,height=11cm,width=11cm}}
\vspace{-0.2cm}
\caption{\small Simulation results for $\Lambda =25 $, $E_{\nu} = 2.5
\times 10^{-4}$, and $E_{\eta} = 2.5\times 10^{-4}$. Same line-style coding as in Fig. \ref{fig:nomagn} for the streamlines.}
\label{fig:highfield}
\end{figure*}

The second simulation is shown in Fig. \ref{fig:highfield}, which
presents the result in the case of a high Elsasser number. This
corresponds to $B_0 = 6.5$ T. The local Elsasser number near the
surface is typically of order of $4.5 \times 10^{-2}$.

In the strong-field case, (see Fig. \ref{fig:highfield}), the system
is essentially dominated by the Lorentz forces everywhere but in diffusive boundary layers. The magnetic field is hardly affected by the
circulation, and keeps essentially its dipolar structure everywhere
within the fluid region. As a result of Ferraro's isorotation law,
the constant angular-velocity contours follow
closely the magnetic field lines. Magnetic connection with the inner
core is strong, and a large region within the outermost closed field line near is forced to
rotate nearly uniformly with angular velocity $\Omega_{\rm in}$. The matching of this uniformly rotating region to the polar regions and to the outer boundary occurs through a diffusive internal shear layer of thickness $\delta_{\parallel}$ 
\begin{equation}
\delta_{\parallel} = \left( \frac{\Enu\Eeta}{\Lambda_{\rm loc}}
\right)^{1/4} \mbox{   ,}
\end{equation}
where $\Lambda_{\rm loc}$ is a local Elsasser number, defined with the local value of the typical amplitude of the field instead of $B_0$.
A more detailed study of this shear layer is outlined in Section \ref{sec:bleq}, and presented by Kleeorin et al. (1997), Dormy, Cardin \& Jault (1998) and Dormy, Cardin \& Soward (2001).

In the polar regions, the field lines are anchored into the differentially rotating convection zone and the latitudinal shear is transmitted to the radiative zone through Ferraro isorotation. The matching of the rotation profile in the polar regions with the outer boundary and the inner core is ensured by the existence of a diffusive Ekman-Hartmann layer with thickness $\delta_{\perp}$ such that 
\begin{equation}
\delta_{\perp} = \left( \frac{\Enu\Eeta}{\Lambda_{\rm loc}}\right)^{1/2}  \mbox{   .}
\end{equation}
This boundary layer is discussed in detail in Section \ref{sec:blradial}. 

The circulation is essentially limited to equatorial regions, with one
strong principal cell following the internal shear layer, and weak
secondary ones. Again, typical flow velocities along the shear layer
are of the order of $r\Ot$ whereas velocities across the layer are of
the order of $\delta_{\parallel} \Ot$. This flow is however too slow
to have any significant effect on the field. In that case, linear
asymptotic analysis can be performed on the system, as shown by
Kleeorin et al. (1997). The toroidal field is mostly limited to
regions of shear (near the poles) with a very small amplitude in the
co-rotating regions. It is worth mentioning that in this case, because
the inner regions rotate almost uniformly, relaxing the rigidity
condition within $r=\rin$ is likely to have little effect on the
solution.

\subsubsection{Intermediate-field case, $\Lambda = 1$}
\label{sec:intfield1}

\begin{figure*}
\centerline{\epsfig{file=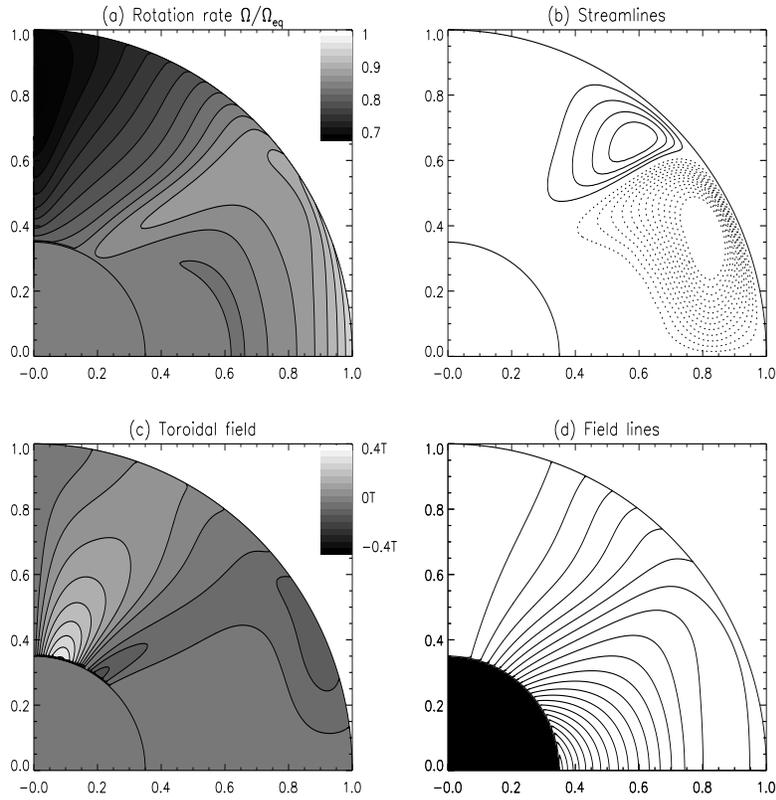,height=11cm,width=11cm}}
\vspace{-0.2cm}
\caption{\small Simulation results for $\Lambda = 1 $, $E_{\nu} =
2.5\times 10^{-4}$, and $E_{\eta} = 2.5\times 10^{-4}$. Same line-style coding as in Fig. \ref{fig:nomagn} for the streamlines.}
\label{fig:intfield1}
\end{figure*}

This third simulation is shown in Fig. \ref{fig:intfield1}, which
presents the result in the case of an intermediate value of the
Elsasser number. This corresponds to $B_0 = 1.3$ T. The local Elsasser
number near the surface is typically of order of $2\times 10^{-3}$.

The intermediate-field case reveals the emergence of two distinct
regions: in the interior the system is dominated by the magnetic
field, and is in a state close to uniform rotation, with a large
region rotating with angular velocity $\Omega_{\rm in}$ (except in a
narrow latitude band around the polar regions). Most of the shear is
confined to a thin layer, the ``tachocline''; 
however, within that shear layer and
especially near the equator, the system can be observed to follow
cylindrical rotation, showing that the system is dominated by Coriolis forces 
rather than Lorentz forces. 

\begin{figure}
\centerline{\epsfig{file=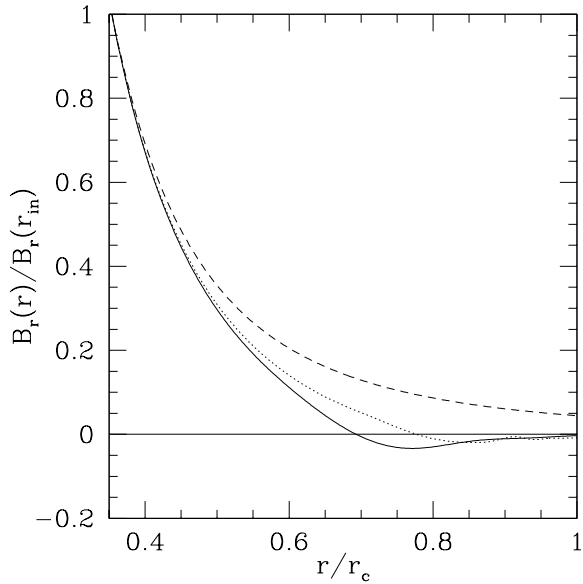,width=8cm}}
\caption{\small Radial component of the magnetic field as a function of
the normalized radius at colatitude $\theta = \pi/3$, 
for the simulation presented in Section \ref{sec:intfield1} (i.e. with the parameter values $\Enu=2.5\times
10^{-4}$, $\Eeta=2.5\times 10^{-4}$ and $\Lambda = 1$) (dotted line),
and in Section \ref{sec:intfield2} (i.e. with the parameter values
$\Enu=6.25\times 10^{-5}$, $\Eeta=6.25\times 10^{-5}$ and $\Lambda =
1$) (solid line). The dashed
line represents the same quantity for a non-rotating system, where the
magnetic field solution is a dipolar field decaying as $r^{-3}$. }
\label{fig:br}
\end{figure}

The following phenomenon is happening. A two-cell circulation with
upwelling in mid-latitudes is driven by Ekman-Hartmann pumping near the
outer boundary, as before. However, in this case the typical circulation velocity ($\ut \sim r\Ot$) is
sufficiently high compared to the Lorentz stresses to advect the
magnetic field in a direction parallel to the outer boundary in
most latitudes, except near the equator, where it is advected
downwards, and near the poles, where the flow is parallel to the
field. As a result of this advection process, the amplitude of the
radial field on the boundary is everywhere reduced, which confines the 
field to the radiative zone and diminishes
the magnetic connection between the interior field and the imposed
latitudinal shear. This can be observed particularly well in Fig. \ref{fig:br}, which shows that the radial component of the field is close to zero in the ``tachocline'' region.
 Moreover the downwards advection of the field in
the equatorial regions reduces the value of the local Elsasser
number (proportional to the square of the amplitude of the field), 
which explains the emergence of the cylindrical rotation in
that area. Flux conservation implies that the
magnetic field strength is correspondingly increased in regions just
below. This magnetic field evacuation by the circulation can also be seen
in Fig. \ref{fig:bt2} which shows the square of the amplitude
of the magnetic field on the equator as a function of radius. 
The plot shows particularly well the two regions:
\begin{enumerate}
\item in the core, the field is hardly perturbed by the differential
rotation imposed on the top, and varies with $r^{-3}$ just as the
field would were the system not rotating (as represented by the dashed
line).
\item the advection of the field by the circulation can easily be seen
near the surface: just below the convection zone, the amplitude of the
field is significantly smaller than the initial dipolar field, and
slightly lower down the amplitude is correspondingly higher, as
required by flux conservation. 
\end{enumerate}
Note that the two regions are separated by a magnetic diffusion layer. Additional simulations for different viscous and magnetic Ekman numbers suggest that the thickness of the layer is, as expected, of order of $\delta_{\parallel}$.

Conversely, the magnetic field keeps the circulation from burrowing
deep into the radiative zone and confines it to a shallow region. This
confinement can be seen in Fig. \ref{fig:intfield1} but is represented
best in Fig. \ref{fig:v2}, which shows the latitudinal component of
the velocity as a function of radius. Note how the circulation is
heavily suppressed below $r=0.9 \rc$.
Finally, one can see that the polar regions, which are unaffected by the circulation, propagate the slow polar rotation at all radii into the radiative zone, as in the high-field case.

\begin{figure*}
\centerline{\epsfig{file=
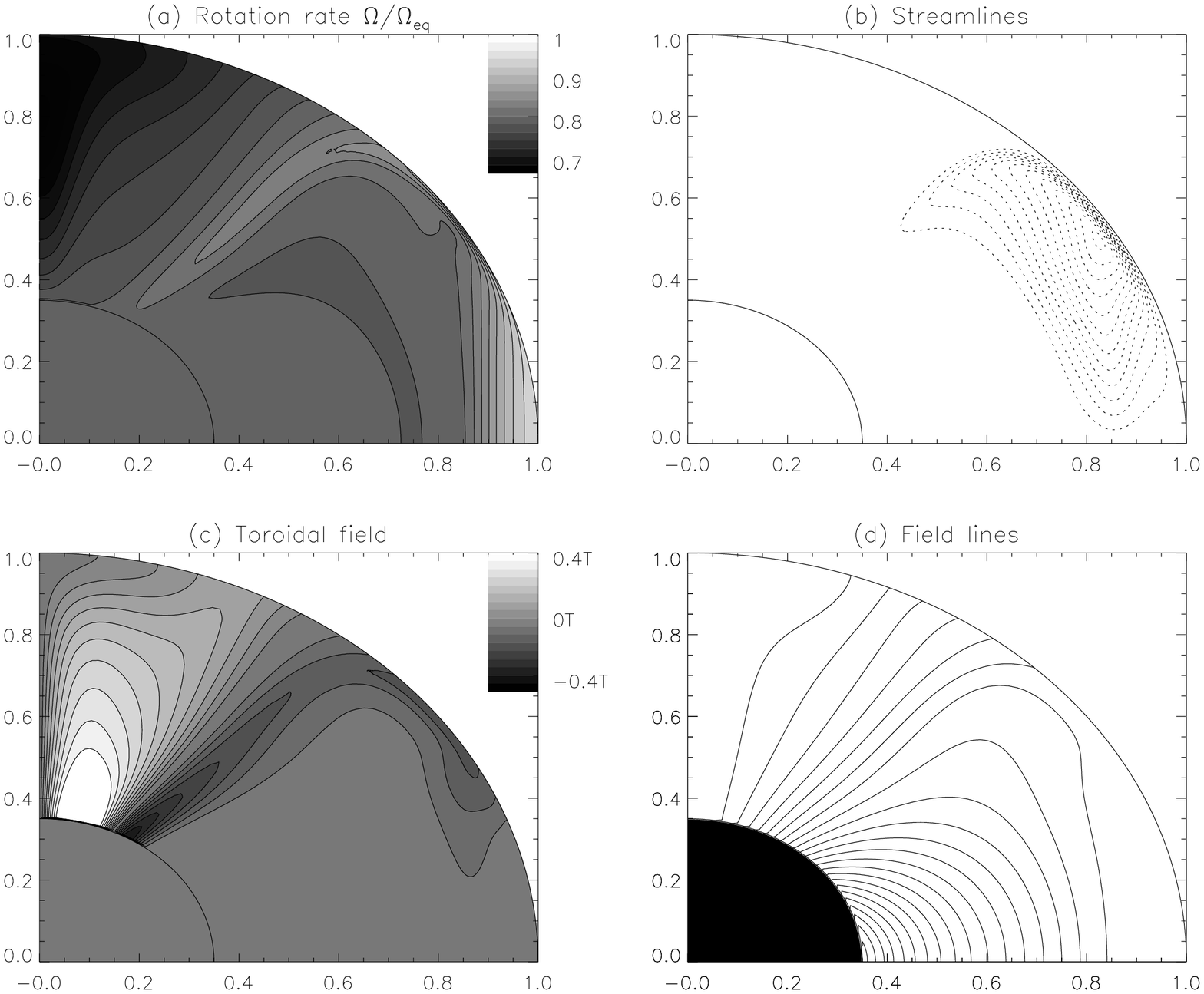,height=11cm,width=11cm}}
\vspace{-0.2cm}
\caption{\small Simulation results for $\Lambda = 1 $, $E_{\nu} =
6.25\times 10^{-5}$, and $E_{\eta} = 6.25\times 10^{-5}$. Same line-style coding as in Fig. \ref{fig:nomagn} for the streamlines.}
\label{fig:intfield2}
\end{figure*}

\subsection{Varying the magnetic diffusivity}
\label{sec:intfield2}

\begin{figure}
\centerline{\epsfig{file=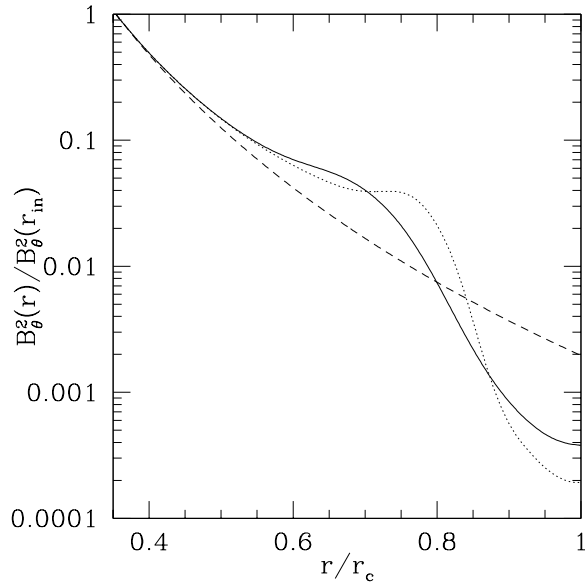,width=8cm}}
\caption{\small Normalized square of the amplitude of the magnetic
field on the equator as a function of normalized radius for the
simulation presented in Section \ref{sec:intfield1} (i.e. with the
parameter values $\Enu=2.5\times 10^{-4}$, $\Eeta=2.5\times 10^{-4}$
and $\Lambda = 1$) (dotted line), and Section \ref{sec:intfield2}
(i.e. with the parameter values $\Enu=6.25\times 10^{-5}$,
$\Eeta=6.25\times 10^{-5}$ and $\Lambda = 1$) (solid line). The dashed
line represents the same quantity for a non-rotating system, where the
magnetic field solution is a dipolar field decaying as $r^{-3}$.}
\label{fig:bt2}
\end{figure}

\begin{figure}
\centerline{\epsfig{file=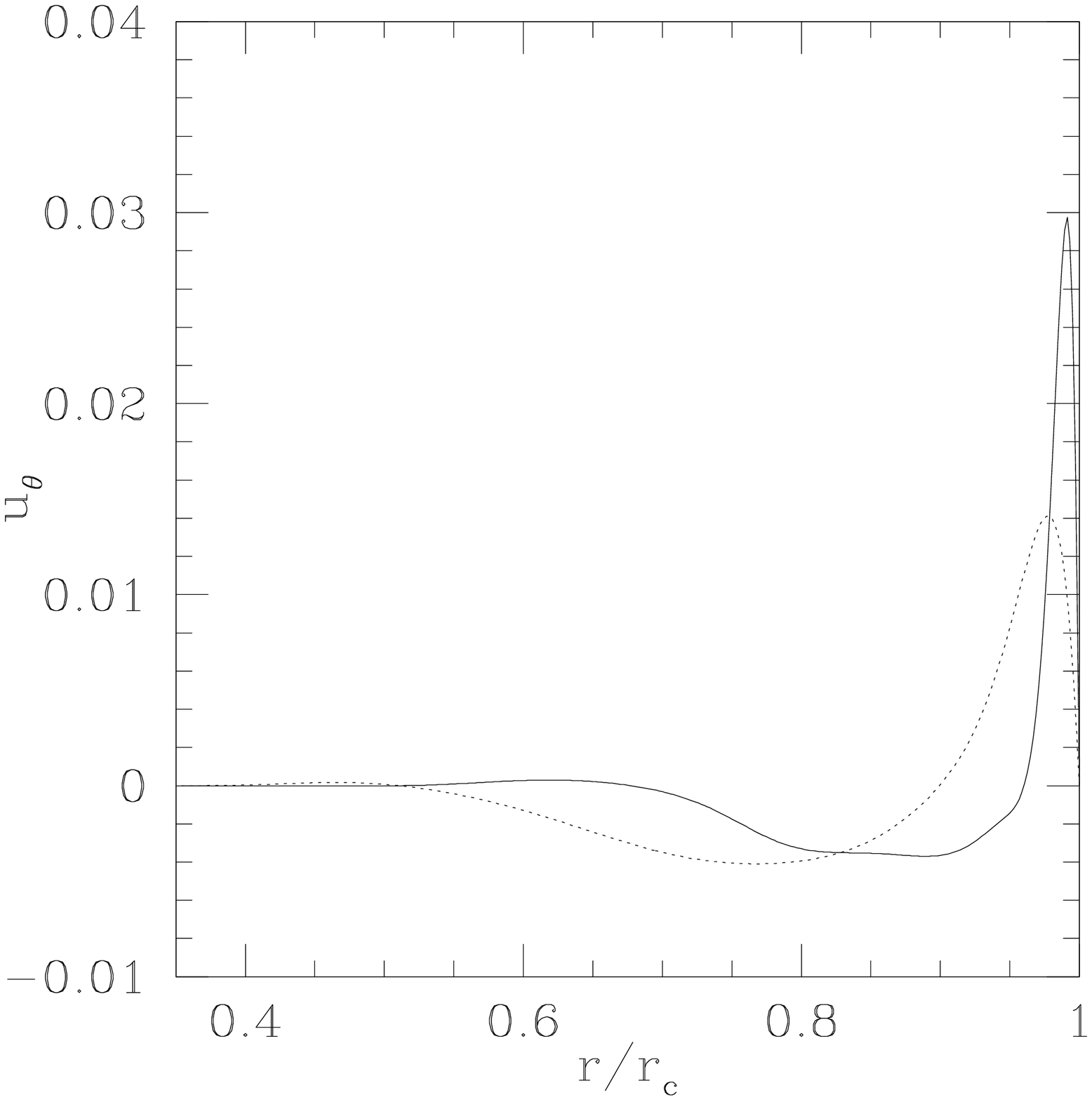,width=8cm}}
\caption{\small Latitudinal component of the velocity as a function of
the normalized radius at colatitude $\theta = \pi/3$, in units of the
azimuthal velocity $\rc\Oc$, for the simulation presented in Section
\ref{sec:intfield1} (i.e. with the parameter values $\Enu=2.5\times
10^{-4}$, $\Eeta=2.5\times 10^{-4}$ and $\Lambda = 1$) (dotted line),
and in Section \ref{sec:intfield2} (i.e. with the parameter values
$\Enu=6.25\times 10^{-5}$, $\Eeta=6.25\times 10^{-5}$ and $\Lambda =
1$) (solid line). }
\label{fig:v2}
\end{figure}

The amplitude of the magnetic field is now set to be that of the intermediate field strength, $B_0 = 1.3T$. When the magnetic diffusivity is decreased by a factor of four, as it is shown in
Fig. \ref{fig:intfield2}, advection of the magnetic field by the
circulation becomes more important compared to diffusion. This has
several consequences:
\begin{enumerate}
\item the fluid is closer to being a perfect fluid, driving the the
system closer to isorotation. A larger volume of fluid in the interior
is rotating nearly uniformly with the interior angular velocity
and the slow rotating polar regions are confined within a smaller
latitude band; 
\item the field confinement and the reduction of magnetic stresses is more efficient, as can be seen in Fig. \ref{fig:br} and Fig. \ref{fig:intfield2}.
\item as can be seen in Fig. \ref{fig:intfield2}  and
Fig. \ref{fig:bt2} the magnetic evacuation in the surface equatorial
regions is also greater, and occurs more abruptly as expected.  The
typical width of the transition layer between the magnetically
dominated interior and the magnetic free tachocline seems to vary as
$(\Enu \Eeta)^{1/4}$, confirming that it is a simple magnetic
diffusion layer of the type analysed in Section \ref{sec:bleq};
\item as a result, the circulation is confined within a smaller volume
by the magnetic field (see Fig. \ref{fig:intfield2} and Fig. \ref{fig:v2}).
\end{enumerate}

\begin{figure}
\centerline{\epsfig{file=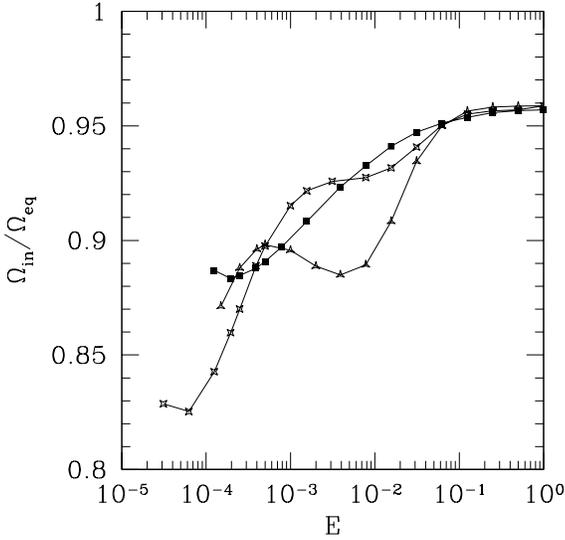,width=8cm}}
\caption{\small Interior angular velocity as a function of the Ekman
number $E=\Eeta = \Enu$ for the low field case (triangles),
intermediate field case (stars) and the high field case (squares) .}
\label{fig:omekm}
\end{figure}

As mentioned previously, quantitative results of the model cannot
provide any serious predictions of the observations (because the
inadequate driving mechanism used for the circulation, because of the
assumption of incompressibility and uniform density, and because of
the gross discrepancy between the real solar diffusivities and the
ones used in these simulations). Nonetheless, the calculated interior
angular velocity provides an interesting point of comparison between
different solutions corresponding to different parameters, and
reflects on the various dynamical processes occuring in the radiative
zone. Figure \ref{fig:omekm} shows the predicted ratio of interior to
equatorial angular velocities as a function of the Ekman number in the
intermediate-field case and high-field case as a function of magnetic
Ekman number. Again, the viscous Ekman number is chosen to have the
same value as the magnetic Ekman number.
The main reason for the initial decrease in $\Omega_{\rm in}$
with decreasing Ekman number is the following: for this range of
magnetic Ekman numbers,  the polar regions are hardly affected by the
shear expulsion, and rotate with a low angular velocity more or less
at all radii. A slowly rotating region, connected to the core via
magnetic stresses, necessarily imposes its slow rotation to the inner
core, with a stronger connection for lower Ekman number. However, as
the Ekman number is decreased past a certain threshold, shear
expulsion from the core slowly starts affecting the polar regions and
the inner core velocity increases again towards the observed
values. Note that the value of the interior angular velocity in the
extremely diffusive case is of $96\%$ of the equatorial angular velocity,
a value which was predicted by Gough (1985).

\section{Comparison with asymptotic analytical analysis}
\label{sec:blanal}

This section focuses on presenting two asymptotic solutions near the
spherical boundaries, in the case where the magnetic field is mostly
perpendicular to the boundary (which occurs near the poles on the
inner boundary) and in the
case where the magnetic field is mostly parallel to the boundary
(which occurs on the outer boundary near the equator). The boundary layer solutions are then compared with the numerical solutions.

\subsection{Analysis of the boundary layer in the polar regions}
\label{sec:blradial}

\subsubsection{Analytical derivation of the boundary layer solution}

When the magnetic field is mostly perpendicular to the boundary, an
Ekman-Hartmann boundary layer develops (see the review by Acheson \&
Hide (1973)). A derivation of the boundary layer analysis is now repeated for the sake of clarity. 
Assuming that the magnetic field is essentially perpendicular
to the boundary with constant amplitude $B_0$, one can write
\begin{equation}
\bB = B_0 \er \mbox{   .}
\end{equation}
Calling $x= r - \rin$, the simplified system of MHD equations 
in the boundary layer on the inner sphere is
\begin{eqnarray}
2\mu \frac{\ptl \psi}{\ptl x} &=& \Lambda \frac{\ptl S}{\ptl x} + \Enu \frac{\ptl^2 L}{\ptl x^2}  \nonumber\\ -
\frac{1}{\rin} \frac{\ptl \psi}{\ptl x} &=& \Eeta \frac{\ptl
b}{\ptl x} \nonumber \\ - 2\mu \frac{\ptl L}{\ptl x}
&=& \rin \Lambda \frac{\ptl^2 b}{\ptl x^2} + \Enu
\frac{\ptl^4 \psi}{\ptl x^4}  \nonumber\\
\frac{\ptl L}{\ptl x} &=& -\Eeta \frac{\ptl^2 S}{\ptl x^2} \mbox{   ,} 
\label{eq:radblin}
\end{eqnarray}
where $\psi$ is the stream function of the fluid flow, such that
\begin{equation}
\st u_{\rm \theta} = \frac{1}{r} \frac{\ptl \psi}{\ptl r} \mbox{    ,   } u_r = \frac{1}{r^2} \frac{\ptl \psi}{\ptl \mu} \mbox{   ,}
\end{equation}
and 
\begin{equation}
L = r\st u_{\phi} \mbox{   ,   } S = r\st B_{\phi} \mbox{   and   } b = \st B_{\theta} \mbox{   .}
\end{equation} 
Grouping these equations yields
\begin{equation}
- 4\mu^2 \Eeta^2 F = \left(\Lambda - \Enu\Eeta \frac{\ptl^2}{\ptl x^2}
  \right)^2 F\mbox{   ,}
\end{equation}
where $F$ represents $\frac{\ptl^2 S}{\ptl r^2}$, $\frac{\ptl^2
b}{\ptl r^2}$, $\frac{\ptl^2 \psi}{\ptl r^2}$ or $\frac{\ptl
L}{\ptl r}$. Looking for a solution of the kind $F \propto e^{\gamma_{\perp}
x}$ yields
\begin{equation}
- 4\mu^2 \Eeta^2 = \left(\Lambda - \Enu\Eeta \gamma_{\perp}^2 \right)^2 \mbox
 {   ,} \end{equation} and in turn, that $\gamma_{\perp} = \pm (\beta \pm
 i\alpha )$ where
\begin{align}
\beta &= \left( \frac{\Lambda + \sqrt{\Lambda^2 + 4\mu^2\Eeta^2}}{2\Enu\Eeta}\right)^{1/2} \simeq \left(\frac{\Lambda}{\Enu\Eeta}\right)^{1/2} \mbox{   ,   } \nonumber 
\\  \alpha &= \left(  \frac{-\Lambda + \sqrt{\Lambda^2 + 4\mu^2\Eeta^2}}{2\Enu\Eeta}\right)^{1/2} \ll \beta \mbox{   .}
\end{align} 
for small enough $\Enu$ and $\Eeta$. The solutions, which must be bounded, can then be written out as
\begin{equation}
Q \simeq  Q_c e^{-\beta x} + Q_0\mbox{   ,}
\end{equation}
where $Q$ is either of the quantities $S$, $b$, $L$ or $\psi$, and
$Q_c$ and $Q_0$ are integration constants. Note that the boundary layer thickness $\delta_{\perp} = 1/\beta$ is, as expected, the same as the one obtained by MacGregor \& Charbonneau (1999) in their open-field calculations.

\subsubsection{Comparison with the numerical solutions}

\begin{figure}
\centerline{\epsfig{file=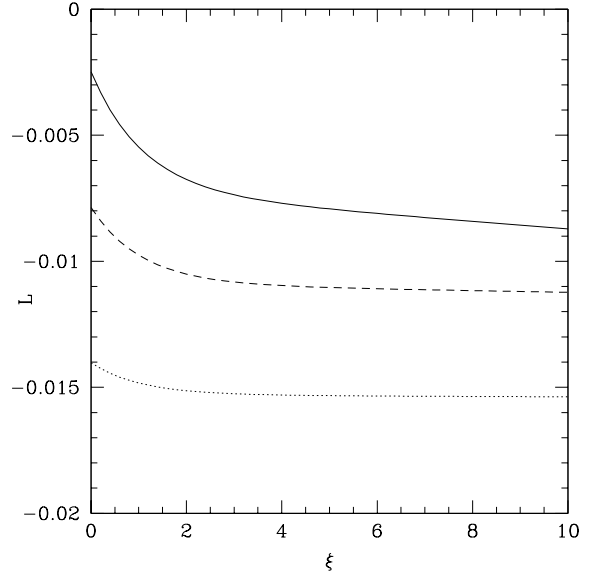,width=8cm}}
\caption{\small Variation with the scaled variable $\xi$ of the angular momentum function $L$ at fixed co-latitude $\theta = \pi/12$ and fixed Elsasser number $\Lambda = 1$, for three different values of the Ekman numbers: $\Enu = \Eeta = 10^{-3}$ (solid line),  $\Enu = \Eeta = 2.5 \times 10^{-4}$ (dashed line) and  $\Enu = \Eeta = 6.25 \times 10^{-5}$ (dotted line).}
\label{fig:blrad}
\end{figure}

In order to compare rigorously the simulations to the analytical
solutions derived above, the matching of the solutions obtained in the
boundary layer to those in the bulk of the fluid should be
performed. However, the solution in the bulk of the fluid, in
particular in the polar regions, is dominated by geometric effects (as
the latitudinal derivatives are not necessarily
negligible) and diffusive effects (as the Ekman numbers used in the
simulations are not small enough to justify neglecting the diffusive
terms); as a result, it is beyond to scope of this analysis to derive
an analytical solution for the solutions in the bulk of the fluid, and
therefore attempt such a matching. However, it is still possible to
check the qualitative behaviour of the solutions in the boundary
layer. Fig. \ref{fig:blrad} shows the angular momentum function $L$ as
a function of the scaled variable $\xi = \beta x = \beta(r-\rin)$, for
three different values of the Ekman numbers. Although the boundary
conditions and asymptotic conditions are different for each case, this
plots illustrates that the angular momentum function $L$ indeed
behaves as $e^{\beta x}$ within the boundary layer (i.e. for values of
$\xi$ below unity). 

\subsection{Analysis of the boundary layer in the equatorial region} 
\label{sec:bleq}

\subsubsection{Analytical derivation of the boundary layer solution}

Near the equator, the asymptotic analysis described in Section
\ref{sec:blradial} breaks down, as the magnetic field is advected by
the circulation into a a direction parallel to the surface. Another
type of boundary layer appears, which is analysed in this section. 

Assuming little
variation in the latitudinal direction, the magnetic field is
approximated by $\bB \approx B_0 \etheta$; also, as only the region near the equator is considered, one can set $\mu \approx 0$ in the following analysis. In that case, and in the
boundary layer only, the MHD equations can be simplified to
\begin{eqnarray}
 \Lambda \frac{B_0}{\rout}\frac{\ptl S}{\ptl \mu} &=&\Enu L''
 \nonumber \\
\frac{B_0}{\rout} \frac{\ptl L}{\ptl \mu}  &=& \Eeta S'' \mbox{   ,}
\end{eqnarray}
where the primes denote differentiation with respect to $r$. These two
equations entirely determine the variation of $L$ and $S$ near the
boundary, and can be combined into
\begin{equation}
\frac{\ptl^2 Q}{\ptl \mu^2} = \rout^2 \frac{\Enu\Eeta}{\Lambda B_0^2} \frac{\ptl^4 Q}{\ptl \zeta^4}\mbox{   ,} 
\end{equation}
where $Q$ is either $L$ or $S$, and where the boundary layer
coordinate $\zeta = \rout-r$ was introduced. On the boundary, the
latitudinal variation of $L$ is given by the boundary conditions, so
that
\begin{equation}
\frac{\ptl^2 L}{\ptl \mu^2} \approx -2(a_2+1)\Omega_{\rm eq} \approx -2(a_2+1)L \mbox{   .}
\end{equation}
So finally, the governing boundary layer equation is
\begin{equation}
-2(a_2+1)L = \rout^2 \frac{\Enu\Eeta}{\Lambda B_0^2} \frac{\ptl^4 L}{\ptl
 \zeta^4} \mbox{   ,}
\end{equation}
and similarly for $S$. The solutions which are bounded as $\zeta
\rightarrow \infty$ are
\begin{equation}
L = L_c(\mu) e^{\gamma_{\parallel} \zeta} \cos (\gamma_{\parallel}\zeta) +
L_s(\mu) e^{\gamma_{\parallel} \zeta} \sin (\gamma_{\parallel}\zeta) \mbox{
,}
\end{equation}
and similarly for $S$, with
\begin{equation}
\gamma_{\parallel} = \left( \frac{\Lambda B^2_0}{2(a_2+1)\rout^2\Eeta\Enu}\right)^{1/4} \propto \left( \frac{\Lambda_{\rm loc}}{\Eeta \Enu}\right)^{1/4} \mbox{  .}
\end{equation}
Note that in the work of R\"udiger \& Kitchatinov (1997), the poloidal field is entirely confined within the radiative zone and, as a result, is everywhere parallel to the surface near the boundary. This type of boundary analysis therefore applies also to their results, with the following modification: as their latitudinal field varies as $(1-\zeta)^{q-1}$ near the boundary, it can be shown that the thickness of the magnetic diffusion layer scales as $\gamma_{\parallel} \propto \left(\Lambda_{\rm loc}/\Eeta \Enu\right)^{1/(2+2q)}$. The case presented here is recovered with $q=1$.

\subsubsection{Comparison with the numerical solutions near the equator}

\begin{figure}
\centerline{\epsfig{file=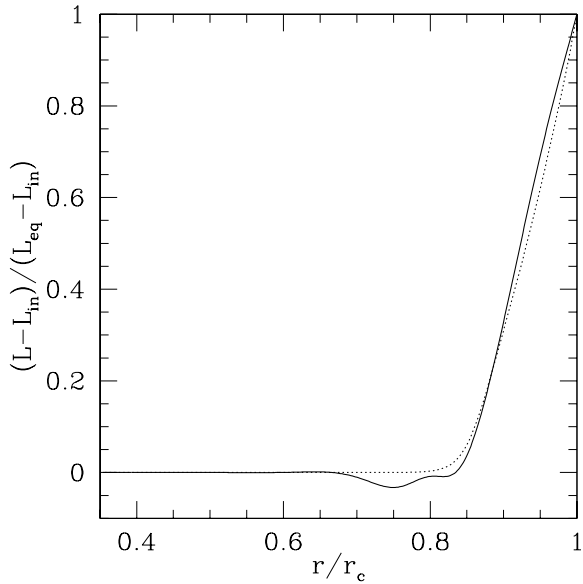,width=8cm}}
\caption{\small Variation with normalized radius of $(L-L_{\rm
c})/(L_{\rm eq}-L_{\rm c})$ on the equator. The solid line is deduced from the numerical solution presented in Section \ref{sec:intfield2} and the dotted line is the exponential solution described in equation (\ref{eq:eqasy})}
\label{fig:bleq}
\end{figure}

Fig. \ref{fig:bleq} presents the variation of $(L-L_{\rm
in})/(L_{\rm eq}-L_{\rm in})$ on the equator as a function of normalized
radius, both for the numerical solution and for the ``analytical prediction'' given by
\begin{equation}
\frac{L-L_{\rm in}}{L_{\rm eq}-L_{\rm in}} \approx e^{\gamma_{\parallel} (r-\rout)}  \mbox{   ,}
\label{eq:eqasy}
\end{equation}
where $L_{\rm in} = \rin^2 \Omega_{\rm in}$.
This simulation corresponds to the parameters $\Eeta = \Enu = 6.25
\times 10^{-5} $ and $\Lambda = 1$. If higher Ekman numbers are
chosen, the asymptotic analysis does not satisfyingly reproduce the
solution. One can see that the general shape of the solution is well
represented by the exponentially decaying solution with boundary layer
width $\delta_{\parallel} = 1/\gamma_{\parallel}$.

The purpose of this section was to determine an asymptotic solution to the equations in the boundary layers near the inner and outer boundaries. This analysis has been used to optimize the mesh-points allocation procedure, as well as to check the behaviour of the numerical solution in the boundaries.

\section{Discussion of the results}
\label{sec:disc}

\subsection{Summary of the results and comparison with previous work}

Section \ref{sec:magn} studied the interaction of a large-scale magnetic
field with fluid motions in the radiative zone when a shear is imposed
through magnetic and viscous stresses by the convection zone. Three
regimes were studied, which depend on the intensity of the magnetic
field. It was shown, by varying the magnetic field strength for
a given set of diffusion parameters ($\Enu,\Eeta)$, that there exists
only a narrow range of parameters $\Lambda$ that leads to a supression
 of the shear in the bulk of the radiative zone. When the
system is dominated by Coriolis forces $(\Lambda \ll 1)$, the shear
extends well into the radiative zone through the Proudman constraint. 
Moreover, two large circulation cells
are driven by Ekman-Hartmann pumping on the boundary, and largely mix
the radiative zone. In the opposite situation ($\Lambda \gg1$)
the system is dominated by Lorentz forces, and the effects of
meridional motion are negligible. In that case, the magnetic field
remains essentially dipolar and the latitudinal shear imposed by the
convection zone extends along the field lines into the radiative
interior. This system was studied in detail by Kleeorin et al. (1997)
and Dormy, Cardin \& Jault, (1998), who showed the existence of an
internal shear layer along the outermost closed field line
``associated with the recirculation of electric currents induced in
the Hartmann layer as the internal Stewartson layer is associated with
the recirculation of meridional flows generated within the Ekman
layers''. The thickness of this internal layer is of order of
$\delta_{\parallel}$ (Dormy, Jault \& Soward, 2001). 
Note that in this strong-field
case it is clear that the internal rotation profile depends
entirely on the structure of the imposed field. This can be deduced from the 
comparative work of MacGregor \& Charbonneau (1999), who show how an
open-field structure, where the field lines are anchored in the
convection zone, results in a differentially rotating radiative
zone, whereas in a confined-field structure the shear can be confined
to a thin tachocline. 

Finally, for $\Lambda \simeq 1$, it is observed that shear exclusion
indeed takes place. In that case, meridional motions significantly
distort the poloidal field structure, to confine it gradually to the
radiative interior as it was initially suggested by Gough \& McIntyre
(1998). This confinement results in the reduction of the magnetic
stresses connecting the radiative interior to the differentially
rotating convection zone, and to the expulsion of the shear by the
interior field to a thin shear layer, as in the simulations first
presented by R\"udiger \& Kitchatinov (1997). However, there exists a
fundamental difference between the shear layer (tachocline) thickness
and position predicted from the R\"udiger \& Kitchatinov and MacGregor
\& Charbonneau (1999) simulations and the ones presented here. In their model
for the confined field, as the field structure is prescribed to the
system and the meridional flows are neglected, the position of the
magnetic diffusion layer is at the boundary and its width is exactly
proportional to  $\delta_{\parallel}(B_0)$, where $B_0$
is the imposed value of the latitudinal field at the boundary. In the
simulations presented here, it is shown that the meridional motions
can affect significantly the strength and the geometry of the field
near the top of the radiative zone through advection processes, which
controls the position and thickness of the magnetic
diffusion layer. This phenomenon is intrinsically linked with the
nonlinear interaction of the field and the flow and can be
studied only through the resolution of the nonlinear MHD equations. The
influence of the flow on the field depends also on the
strength and on the geometry of the flow, which in turn are
intrinsically related to the mechanism which drives the flow. These comments
naturally build up to the model proposed by Gough \& McIntyre (1998). 

One of the principal features of the Gough \& McIntyre model is their
suggestion that the magnetic diffusion layer could be situated below a
virtually magnetic-free tachocline (by magnetic-free, they refer more
to a system where the influence of the magnetic field in the dynamical
equilibrium is negligible than to a system that has no magnetic field
at all). This structure would result from the almost complete confinement 
of the magnetic field within the radiative zone by the meridional motions (except perhaps in a localized upwelling region in mid-latitudes).
Although the simulations above cannot reproduce quantitatively the
dynamical balance proposed by Gough \& McIntyre (as the circulation in
this paper is artificially driven by Ekman-Hartmann pumping instead of
thermal imbalance), these qualitative features can easily be
identified in Fig. \ref{fig:br} and Fig. \ref{fig:bt2} . The system
seems indeed to lead naturally to a magnetically dominated interior
and a rotationally dominated tachocline separated by a magnetic
diffusion layer of a thickness that is observed to vary as
$\delta_{\parallel}$.

\subsection{Comparison with observations}

The next step of this analysis consists in comparing the numerical
simulations with the observations. In this comparison process, it is essential to keep in mind two essential discrepancies between the model and the real sun: the driving mechanism for the meridional flow in these simulations is artificial, so that the typical flow velocities may be erroneous, and the typical Ekman numbers of the simulations are several orders
of magnitude larger than in the sun. In the region of the tachocline,
assuming that the flow is not turbulent, the magnetic and viscous
diffusion coefficients are of order of $\nu = 10$cm$^2$s$^{-1}$ and
$\eta = 2\times 10^{3}$cm$^2$s$^{-1}$, which implies that
\begin{equation}
\Enu \simeq 10^{-15} \mbox{   and   } \Eeta \simeq 2\times 10^{-13} \mbox{   .}
\end{equation}
The main consequence of these discrepancies is that although the
principal features of the interaction between fluid motions and
large-scale fields in the sun can be studied through this method, it
cannot provide reliable quantitative estimates.

Despite this obvious shortfall, some aspects of the observations can
be reproduced in the simulations.
\begin{enumerate}
\item For sufficiently low magnetic diffusivity, a large region of the
radiative zone  is forced to rotate uniformly with the angular
velocity of the core; such a  uniform angular-velocity profile in the
radiative interior is indicated clearly by the observations. The
simulations suggest that polar regions seem to rotate with a low
angular velocity down to a latitude of about $50^{\circ}$ for  $\Enu =
\Eeta = 2.5 \times 10^{-4}$, and down to latitudes of about
$60^{\circ}$ for $\Enu = \Eeta = 6.25 \times 10^{-5}$: the polar shear
is gradually confined to higher and higher latitudes as the magnetic Ekman
number is reduced. Helioseismic inversions still have too low a
resolution to  provide any reliable observations of the polar regions;
this numerical  model however predicts more slowly rotating fluid in
the polar regions deep  within the radiative zone, which is a natural
feature of the dipolar field configuration (R\"udiger \& Kitchatinov
1997, Gough \& McIntyre 1998, MacGregor \& Charbonneau 1999). Since the
most likely configuration for a fossil field is dipolar, this polar
feature is a robust prediction of this family of models.
\item A thin boundary layer appears in which most of the shear is
contained, particularly in the equatorial regions. The thickness of
this shear layer is much larger than the observed thickness of the
tachocline, a discrepancy which is again simply related to the high
diffusivities adopted here. It is interesting to note, however, that the
position of the magnetic diffusion layer predicted by the
simulations is  deeper in the equatorial regions than at higher
latitudes (which is due to the downward advection of the field near the equator); this could be related to observations of the prolateness of
the tachocline (Charbonneau et al, 1999).
\item A two-cell meridional circulation is driven below the convection
zone by  Ekman-Hartmann pumping; this circulation is confined by the
large-scale poloidal field to the tachocline region. This result
validates the analysis of the sound-speed profile performed by Elliott
\& Gough (1999) and also relates to  the upper limits in the depth of
the tachocline mixing suggested by  observations of the abundances of
light elements in the convection zone (Brun, Turck-Chi\`eze \& Zahn
(1999)).  Again, the simulated depth of penetration of  the circulation
is much larger than suggested by the observations, a  discrepancy
which is due to the large diffusivities used for the numerical
analysis, and to the assumption of incompressibility: comparison
between high- and low-diffusivity cases indeed show a significant
reduction of the mixed-layer depth for lower Ekman numbers, and when
the background stratification is taken into account, it is observed to
 impede radial motions dramatically(see subsequent work).
\end{enumerate}

Finally, as mentioned in Section \ref{sec:intfield2}, a useful point
of comparison between different theoretical models and between models and the
observations is the value of the interior angular velocity $\Omega_{\rm
in}$. Gough (1985) showed that for an incompressible tachocline
controlled by viscous effects, $\Oin = 0.96 \Oeq$. 
MacGregor \& Charbonneau (1999) presented simulations in a
confined field configuration which show a virtually uniform rotation
profile for the radiative zone with $\Oin = 0.97\Oeq$. The discrepancy
with the observed value ($\Oc = 0.93 \Oeq$) is significant, and
comparison with the Gough \& McIntyre model shows that either
meridional motions and the effects of heat transport and
stratification are dominant in the dynamics of the tachocline, or that
such a class of models cannot reproduce the observations and that very
different dynamics are in play (as in the turbulent stress model
proposed by Spiegel \& Zahn, 1992, or the nonlinear 
development of MHD instabilities presented by Cally, 2001). 
The prediction for the angular velocity presented in Fig. \ref{fig:omekm} are not conclusive
when comparing them with observations, as any discrepancy could be
attributed to the large diffusivities used in the simulations.

\subsection{Discussion of the model}
\label{sec:discmod}

The ultimate aim of this work is to develop a self-consistent dynamical
model of the tachocline which can be used to explain the large range
of observations available. In order to do so, the idea proposed by
Gough \& McIntyre is gradually implemented into a numerically solvable
nonlinear MHD model. Although previous work on those lines (R\"udiger
\& Kitchatinov, 1997, MacGregor \& Charbonneau, 1999) has already
investigated some of the aspects of the interaction of a large-scale
field and differential rotation, two essential ingredients to the
model remained to be studied carefully:
\begin{enumerate}
\item the nonlinear interaction of a meridional circulation and the imposed magnetic field
\item the baroclinic driving of meridional motions and the effects of compressibility and stratification. 
\end{enumerate}
This first paper focuses on the first point only by artificially
driving a meridional flow with Ekman-Hartmann pumping on the
boundaries. It is found that this artificial system reproduces
qualitatively (but not quantitatively) most of the aspects of the
Gough \& McIntyre model, and of the observations. The good qualitative
agreement with the Gough \& McIntyre model is perhaps surprising, but
is simply due to the qualitative similarities between the
Ekman-Hartmann flow and the baroclinicly driven flow, in particular
with respect to the two-cell structure with upwelling in
mid-latitudes. In both cases, the quadrupolar structure of the flow is
linked with the transition in mid-latitudes from a negative radial
shear to a positive radial shear. When Ekman-Hartmann layers are
present on the outer boundary, they naturally drive 
poleward motions at high latitudes and equatorward motions at low latitudes, 
with a return flow upwelling in mid-latitudes. In the sun,
thermal-wind balance across the tachocline 
is shown to lead to temperature fluctuations
with a minimum in mid-latitudes, which also results in an upwelling
in this region (Spiegel \& Zahn, 1992). 

\begin{figure}
\centerline{\epsfig{file=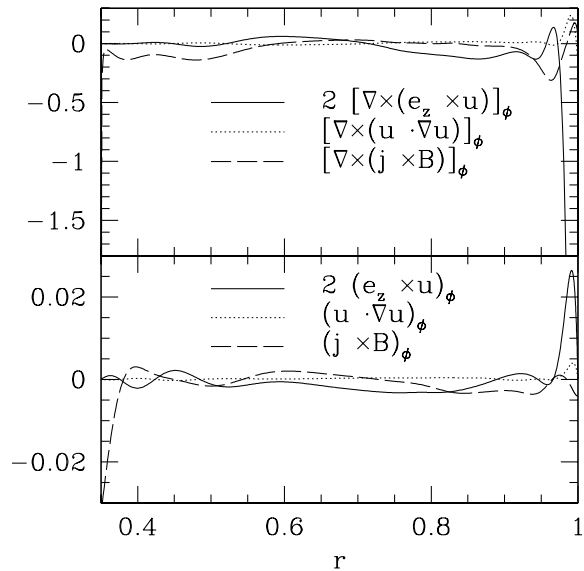,width=8cm}}
\caption{\small Lower panel: comparison between the advection term, the Coriolis force and the Lorentz force at co-latitude $\theta = 60^{\circ}$ in units of $\rc\Oc^2$, for the simulations presented in Section \ref{sec:intfield2}. Upper panel: comparison between the vorticity advection term, the background vorticity advection and the magnetic stresses at co-latitude $\theta = 60^{\circ}$ in units of $\Oc^2$.} 
\label{fig:nonlin}
\end{figure}

Another important point concerns one of the principal assumptions of this numerical model: the linearization of the problem with respect to the advection of momentum by the meridional flow. In order to verify that $(\bu \cdot \grad) \bu$ can indeed be
neglected, it is compared in Fig. \ref{fig:nonlin} to the linear term
$2\ez \times \bu$ as well as the Lorentz force $\bj \times \bB$, for the simulation presented in Section \ref{sec:intfield2} and in Fig. \ref{fig:intfield2}.
 As assumed, the nonlinear advection term is everywhere much smaller than
the Coriolis term, except perhaps in the Ekman-Hartmann layers near $r=\rin$
where both Coriolis and advection term vanish (the system is
then dominated by the magneto-viscous balance). It is therefore
justified to neglect it. A similar comparison can be performed (see Fig. \ref{fig:nonlin})
between the vorticity advection term $\curl (\bu \cdot \grad \bu)$,
the background vorticity advection term $2\curl (\ez\times \bu)$ and
the Lorentz stresses $\curl(\bj\times\bB)$; the conclusion is the
same. 
 \\
\\
Finally, a few more points must also be raised.
\begin{enumerate}
\item As mentioned previously, the diffusivities used in the
simulations are far higher than in the sun. The principal effect of
high diffusivities is to reduce the interconnection between
the field and the flow which can significantly change the balance of forces within the fluid region. This discrepancy is forced by numerical solvability and can be reduced at great cpu expense only.
\item The boundary conditions on the magnetic field are found to have significant influence on the problem (MacGregor \& Charbonneau 1999). Although the systematic study of the effects of boundary conditions on the system has not been discussed in this paper, it is illustrated, for example, in the comparison between this work and the results of Dormy, Cardin \& Jault (1998) who study (in a geophysical application) the effects of an insulating outer boundary; as a result, the structure of the toroidal field in the interior is altered significantly.
\item The boundary conditions near the top boundary (which is the base
of the convection zone) poorly represent the dynamical interaction
between the flow and field in the tachocline and the convection zone: 
the convection zone, and to some extent the overshoot region
below, is the seat of fully developed turbulence. It does not behave as
an impermeable wall to the circulation in the tachocline, or as a uniformly conducting medium to the magnetic field. Moreover, it is for
example widely believed that the interface of the tachocline and the
convection zone is the seat of the solar dynamo, which is a non-steady
magnetic structure oscillating with a 22 yr period. 
How can this structure be matched with
the assumed steady, nearly dipolar field of the interior? The problem was avoided in this work by assuming that the upper boundary of the tachocline lies safely below the region of influence of the dynamo (see Garaud, 1999) and the overshoot region, but this is clearly an over-simplification. 
\item This also raises one of the major issues of the dynamics
of the tachocline: to what extent can a steady model represent the
system? Firstly, how can the interior field be represented by a steady field when part of it diffuses, or is advected away into the convection zone? In the Gough \& McIntyre model, the diffusion
of the magnetic field out of the radiative zone is impeded by the
confining action of the circulation on the poloidal field, which
supports the steady-state assumption. However, the upwelling region
discussed by Gough \& McIntyre (and observed in
Fig. \ref{fig:intfield1}) drags the poloidal magnetic field 
into the convection zone, where it could occasionally reconnect. This phenomenon
is intrinsically non-steady, although possibly
quasi-periodic (as no magnetic flux is lost through reconnection). Secondly, the hydrodynamical stability of the solution
to various instabilities (and in particular to non-axisymmetric perturbations) 
should be assessed. It was shown by Garaud (2001a) that two-dimensional hydrodynamical instabilities are unlikely to be strong enough to have a significant effect on angular-momentum redistribution in the tachocline. However, the stability of the
shear flow tends to be upset by the presence of a magnetic field, as
suggested by Gilman \& Fox (1997), Dikpati \& Gilman (1999) and more recently by Cally (2001). In a stratified fluid, the buoyancy
of toroidal flux tubes may lead to a magneto-convective
instability. Finally, Alfv\'en waves can propagate along the magnetic
field and lead to the presence of either ``localized'' oscillations,
or global torsional oscillations. The combined effect of these instabilities and waves on the flow may be important in redistributing angular momentum within the radiative zone and across the tachocline. This has not been taken into account in this model.
\end{enumerate}

In future work each of these problems must be addressed. The gradual
convergence of the system towards lower and lower diffusivities is
``merely'' a computational challenge, which can in principle be
solved. There is hope that for low enough diffusivities the system
reaches an asymptotic state which is more-or-less independent of the
Ekman parameters. Asymptotic analyses, following the steps of Kleeorin
et al. (1998) for instance, may provide a way of by-passing boundary
layers in the flow by providing jump-conditions across the
layers. This would greatly help reducing some of the numerical
difficulties. However, looking at the importance of geometric or
nonlinear effects in the system, it is unlikely that an asymptotic
analysis could provide a full solution of the problem. The problem of
the representation of the dynamical interaction of the tachocline and
the convection zone through adequate boundary conditions is a strong
theoretical challenge, which will be looked into in the future. The
most obvious route is through the inclusion of the convection zone in
the computational domain, and the ad-hoc prescription of Reynolds
stresses which would lead to the observed rotation profile. But
comparing the relative importance of all the points mentionned in this
section, it is clear that the next step towards improving the model is
through the inclusion of the effects of compressibility, energy
transport and stratification on the dynamics of the system, which will
allow a correct quantitative representation of the baroclinic driving of
the meridional motions as well as the effects of the strong
stratification in the radiative zone. This is work which will be
presented in a subsequent paper.

\section{Conclusion}

This paper presents a numerical analysis of the nonlinear 
interaction between a primordial field and large-scale fluid flows 
in the solar radiative zone when a latitudinal shear is imposed from the 
convection zone through a combination of magnetic and viscous stresses. 

Within the scope of some simplifying assumptions (axisymetry,
steady-state, incompressibility), the presence of a large-scale field
in the radiative zone was shown to lead to three dynamical regimes, 
which depend essentially on the strength of the
underlying field.  When the Elsasser number is low, the flow is
dominated by the Proudman constraint, which enforces a rotation
profile roughly constant on cylindrical surfaces throughout most of the radiative zone. When the Elsasser
number is high, the magnetic field propagates the shear to a large part 
of the interior according to the Ferraro isorotation law, and only within the outermost closed field line does one observe uniform rotation.  
When the Elsasser number is of order of unity, the magnetic field is weak
enough to be advected by meridional motions driven by Ekman-Hartmann 
pumping on the boundary. This advection process stretches the field along the boundary (thereby reducing the magnetic stresses connecting the convection zone and the radiative zone) and pushes the poloidal field deeper into
the interior. As a result, the shear is confined to some shallow layer below the convection zone, which can be likened to the tachocline. Conversely, the meridional circulation is kept from
burrowing into the radiative zone by the accumulation of magnetic flux
below the tachocline.

The structure of the solution in the intermediate-field case is in
qualitative agreement with the angular-velocity observations, as well as the predictions of the Gough \& McIntyre (1998) model. 
The simulations also show the presence of efficient mixing in the
tachocline which would influence the abundances of chemical elements
in that region; this can also be detected observationally (Elliott \&
Gough, 1999, Brun, Turck-Chi\`eze \& Zahn, 1999).

As expected, the quantitative agreement between the model and the
observations is poor; this is due on the one hand to the large
diffusivities used in the simulations, and on the other hand to some
oversimplification of the dynamics involved (through the boundary
conditions chosen, and also the assumption of incompressibility, which 
drive a meridional flow quantitatively different from what one could 
expect in the tachcoline). Both
of these problems will be addressed in future work; in particular, the
effects of compressibility will be studied in a following paper.

In conclusion, although the dynamical behaviour of the sun is likely to
be far more complex than that of the simulations presented in this
paper, it is clear that these simulations can grasp, and
explain, some of the {\it fundamental} aspects of the dynamics of the
observed solar differential rotation. Moreover, they lay a strong
basis for future improvements of the models along the lines described
above.

\section*{Acknowledgments}

This work was funded by New Hall, PPARC and the Isaac Newton Studentship at various stages of its completion. None of this work would have been possible without the constant encouragements and tremendous insight of Professor D. O. Gough. I thank Professors M. R. E. Proctor, N. O. Weiss, and J.-P. Zahn, as well as an anonymous referee, for reading this work and for their very constructive criticisms which greatly helped improve on the clarity of this paper.

\section*{Appendix: Asymptotic solution in the non-magnetic case}

Following the work of Proudman (1956), the equations
\begin{eqnarray}
(\ez \times \bu)_{\phi} = \Enu (\grad^2 \bu)_{\phi} \mbox{    ,} \nonumber \\
\left[ \curl (\ez \times \bu) \right]_{\phi} = \Enu (\grad^2 \bomega)_{\phi} \mbox{    ,}  
\label{eq:31}
\end{eqnarray}
are first solved in the main
 body of the fluid, then successively near the bottom and top boundaries.
 In the main body of the fluid, the viscous stresses are negligible, so that $\bu$ is independent of $z$, and the solutions are
\begin{equation}
\psi=\psi_0(s) \mbox{   and   } \chi = \chi_0(s) \mbox{   .}
\end{equation}
where $\chi$ is defined such that
\begin{equation}
\up = \frac{\chi}{r\st} \mbox{   .}
\end{equation}
In order to solve the problem near the lower boundary, Proudman introduces
 the stretched variable $\zeta$ such that
\begin{equation}
\zeta = (r-\rin)\Enu^{-1/2} \cos^{1/2}\theta \mbox{   .}
\label{eq:boundlay}
\end{equation}
This recognizes the presence of a boundary layer with thickness 
$\delta_{\nu} = \Enu^{1/2} \cos^{-1/2}\theta$. The equations (\ref{eq:31}) become, to zeroth order in $\Enu^{1/2}$
\begin{eqnarray}
\frac{\ptl^5 \psi}{\ptl \zeta^5} &=& -4 \frac{\ptl \psi}{\ptl \zeta} 
\label{eq:tea3} \nonumber\\
\frac{\ptl^5 \chi}{\ptl \zeta^5} &=& -4 \frac{\ptl \chi}{\ptl \zeta} \mbox{   .}
\label{eq:tea4}
\end{eqnarray}
Note that these approximations are not valid near the poles, where
latitudinal derivatives may become important, and near the equator,
where the thickness of the boundary layer $\delta$ diverges. 

Assuming that the whole system is rotating with angular velocity $\Omega_{\rm in}$ so that effectively $\Oc = \Omega_{\rm in}$ (this only calls for a slight re-definition of the Ekman number), matching with the boundary conditions at the bottom boundary
requires that $\chi \rightarrow 0$ as $r\rightarrow 1$, or $\zeta
\rightarrow 0$. Moreover, the impermeable boundary condition requires
that $\psi = 0$ on the boundary, as well as $\ptl \psi/\ptl
\zeta=0$. The solution to equations (\ref{eq:tea3}) which fulfills all these boundary conditions, and
which is bounded as $\zeta \rightarrow \infty$ is
\begin{eqnarray}
\psi(\zeta,\theta) &=& \psi_1(\theta) \left(1-e^{-\zeta}(\cos\zeta + \sin\zeta) \right) \nonumber \\
\chi(\zeta,\theta) &=& \chi_1(\theta) \left(1-e^{-\zeta}\cos\zeta\right) \mbox{   ,}
\end{eqnarray}
where $\psi_1(\theta)$ remains to be determined, and
\begin{equation}
\chi_1(\theta) = 2 \Enu^{-1/2} \cos^{1/2}\theta  \psi_1(\theta) \mbox{   .} 
\label{eq:teabc1}
\end{equation}
As $\zeta \rightarrow \infty$, these functions must match onto the
solution obtained previously for the main body of the fluid region so that
is then easy to see that one must have
\begin{equation}
\chi_1(\theta) = \chi_0(\rin\st) \mbox{   and   } \psi_1(\theta) = \psi_0(\rin\st) \mbox{   .}
\end{equation}
This result can be combined with equation (\ref{eq:teabc1}) and yields
the matching condition
\begin{equation}
\chi_0(\rin\st) = 2 \Enu^{-1/2} \cos^{1/2}\theta \psi_0(\rin\st) \mbox{   .}
\label{eq:teabc2}
\end{equation}
In order to study the boundary layer near the top boundary, another
stretched variable is introduced:
\begin{equation}
\xi = (\rout-r) \Enu^{-1/2} \cos^{1/2}\theta \mbox{   .}
\end{equation}
The scaled equations are the same as before (cf equations (\ref{eq:tea4})); the boundary conditions for the
stream function are also the same as for the lower boundary when $\chi
\rightarrow 0$, but the differential rotation must now match onto that
of the convection zone, so that
\begin{equation}
\chi(r=\rout, \theta) = \rout^2 \s2t \tilde{\Omega}_{\rm cz}(\theta)\mbox{   ,} \end{equation}
where
\begin{equation}
\tilde{\Omega}_{\rm cz}(\theta) = \Omega_{\rm eq} (1- a_2\c2t -a_4\cos^4\theta) -\Omega_{\rm in} \mbox{   .} 
\end{equation}
The solutions to equations (\ref{eq:tea3}) which
fulfill these conditions are
\begin{eqnarray}
\psi(\xi,\theta) &=& \psi_2(\theta)\left(1-e^{-\xi}(\cos\xi + \sin\xi)\right) \mbox{   ,} \nonumber\\
\chi(\xi,\theta) &=& \chi_2(\theta) + 2 \Enu^{-1/2} \cos^{1/2} \theta \psi_2(\theta) e^{-\xi} \cos\xi \mbox{   , }
\end{eqnarray}
with
\begin{equation}
\chi_2(\theta) = \rout^2 \s2t \tilde{\Omega}_{\rm cz}(\theta) - 2\Enu^{-1/2} \cos^{1/2}\theta \psi_2(\theta) \mbox{   .}
\end{equation}
As before, matching with the solution in the main body of the
fluid implies that
\begin{equation}
\psi_2(\theta) = \psi_0(\rout\st) \mbox{   and   } \chi_2(\theta) = \chi_0(\rout\st) \mbox{   ,}
\end{equation}
so that
\begin{equation}
\chi_0(\rout\st) = \rout^2 \s2t \tilde{\Omega}_{\rm cz}(\theta) - 2\Enu^{-1/2} \cos^{1/2}\theta \psi_0(\rout \sin\theta) \mbox{   .}
\label{eq:teabc3}
\end{equation}
Since $\psi_0$ and $\chi_0$ are functions of $s$ only, the two
matching conditions given by equations (\ref{eq:teabc2}) and
(\ref{eq:teabc3}) can also be rewritten as
\begin{align}
\chi_0(s) &= 2 \Enu^{-1/2} \left(1-(s/\rin)^2\right)^{1/4} \psi_0(s) \mbox{   ,}\nonumber\\
\chi_0(s) &= s^2 \Omega'_{\rm cz}(s) - 2\Enu^{-1/2} \left(1-(s/\rout)^2\right)^{1/4} \psi_0(s) \mbox{   ,}
\end{align}
where
\begin{eqnarray}
\Omega'_{\rm cz}(s) &=& \Omega_{\rm eq} \left[1 -  a_2\left(1-(s/\rout)^2\right) \right.  \nonumber \\ && \left. - a_4\left(1-(s/\rout)^2\right)^2 \right] -\Omega_{\rm in} \mbox{   ,}
\label{eq:om'cz}
\end{eqnarray}
which can now be solved uniquely as
\begin{align}
\psi_0(s) =& \frac{\Enu^{1/2}}{2} \frac{s^2 \Omega'_{\rm cz}(s)}{ \left(1-(s/\rin)^2\right)^{1/4} + \left(1-(s/\rout)^2\right)^{1/4}} \nonumber\\
\chi_0(s) =& \frac{s^2 \left(1-(s/\rin)^2\right)^{1/4} \Omega'_{\rm cz}(s)}{ \left(1-(s/\rin)^2\right)^{1/4} + \left(1-(s/\rout)^2\right)^{1/4}} \mbox{   .}
\end{align}  
The flow within two spheres is now known analytically everywhere.

Assuming that the sun is in equilibrium, the total torque applied by
the convection zone on the radiative interior should be equal to that
exerted by the solar wind on the convection zone. Since that torque is
extremely small, it is assumed to be null as a first approximation,
which is equivalent to requiring that the sun be in a steady
state. This condition determines the interior angular velocity 
$\Omega_{\rm in}$
uniquely. 

In the non-magnetic case, the torques applied by the tacho\-cline
onto the radiative zone are purely viscous. As a result, the
steady-state condition can be rewritten as
\begin{equation}
T_{\nu}(r=1)= 2\pi \nu \int_0^{\pi/2} \left[ r^3 \s2t \frac{\ptl \Omega}{\ptl r}\right]_{r=1} \st \dd \theta   = 0 \mbox{   ,}
\end{equation}
where $T_{\nu}$ is the total viscous torque and $\Omega = \Omega_{\rm in} + (\chi/r^2\s2t)$ is the total angular velocity at the base of the fluid region. Using the results derived previously, this condition can be rewritten as:
\begin{equation}
\frac{\Omega_{\rm in}}{\Omega_{\rm eq} } = \frac{\int_0^{\pi/2} F(\theta) D(\sin \theta) \dd \theta}{\int_0^{\pi/2} F(\theta) \dd \theta}
\label{eq:omA} \mbox{   ,}
\end{equation}
where 
\begin{equation}
F(\theta) = \frac{\sin^3\theta  \cos\theta}{\cos^{1/2}\theta + \left(1-\s2t\rin^2/\rout^2 \right)^{1/4}} 
\end{equation}
and
\begin{equation}
D(s) = 1-a_2 \left(1-(s^2/\rout^2)\right)-a_4\left(1-(s^2/\rout^2)\right)^2 \mbox{   .}
\label{eq:ds} 
\end{equation}

\end{document}